\documentclass[12pt]{article}

\usepackage{amsmath}
\allowdisplaybreaks
\usepackage{epsf}
\usepackage{amssymb}
\usepackage{bm}
\usepackage{mathrsfs}
\usepackage{wick}

\begin{document}

\baselineskip=16.75pt plus 0.2pt minus 0.1pt

\makeatletter
\@addtoreset{equation}{section}
\renewcommand{\theequation}{\thesection.\arabic{equation}}
\newcommand{\calA}{{\cal A}}
\newcommand{\calB}{{\cal B}}
\newcommand{\calC}{{\cal C}}
\newcommand{\calE}{{\cal E}}
\newcommand{\calP}{{\cal P}}
\newcommand{\calM}{{\cal M}}
\newcommand{\calN}{{\cal N}}
\newcommand{\calV}{{\cal V}}
\newcommand{\calK}{{\cal K}}
\newcommand{\calF}{{\cal F}}
\newcommand{\calG}{{\cal G}}
\newcommand{\calH}{{\cal H}}
\newcommand{\calT}{{\cal T}}
\newcommand{\calU}{{\cal U}}
\newcommand{\calY}{{\cal Y}}
\newcommand{\calW}{{\cal W}}
\newcommand{\calL}{{\cal L}}
\newcommand{\calD}{{\cal D}}
\newcommand{\calO}{{\cal O}}
\newcommand{\calI}{{\cal I}}
\newcommand{\calQ}{{\cal Q}}
\newcommand{\calS}{{\cal S}}
\newcommand{\QB}{{\cal Q}_\text{B}}
\newcommand{\spQ}{\mbox{\boldmath $\QB$}}
\newcommand{\nn}{\nonumber}
\newcommand{\deriv}[2]{\frac{d #1}{d#2}}
\newcommand{\eps}{\varepsilon}
\newcommand{\ro}{\biggr|_{r=0}}
\newcommand{\ds}{\displaystyle}
\newcommand{\Fe}{F_{\eps}}
\newcommand{\Ke}{K_{\eps}}
\newcommand{\Ue}{U_{\eps }}
\newcommand{\ze}{\text{z}_{\eps}}
\newcommand{\asn}{\bigger|}
\newcommand{\asne}{\bigger|_{\eps}}
\newcommand{\ePsi}{_{\eps}\Psi}
\newcommand{\Psie}{\Psi_{\eps}}
\newcommand{\ppsi}{\Psi_{p\eps}}
\newcommand{\vev}[2]{\langle{ #1}\rangle_{\ell= #2}}
\newcommand{\CR}[2]{\left[#1,#2\right]}
\newcommand{\ACR}[2]{\left\{#1,#2\right\}}

\newcommand{\tr}{\mathop{\rm tr}}
\newcommand{\Tr}{\mathop{\rm Tr}}
\newcommand{\wN}{N}
\newcommand{\swN}{\calN}
\newcommand{\p}{\partial}
\newcommand{\wh}[1]{\widehat{#1}}
\newcommand{\abs}[1]{\left| #1\right|}
\newcommand{\KeR}{$\Ke$-regularization}
\newcommand{\CSFTwN}{winding number}
\newcommand{\Utv}{U_\text{tv}}
\newcommand{\wt}[1]{\widetilde{#1}}
\newcommand{\VEV}[1]{\left\langle #1\right\rangle}
\newcommand{\Drv}[2]{\frac{\p #1}{\p #2}}
\newcommand{\hanaL}{\mathcal{L}}
\newcommand{\hanaLhat}{\wh{\mathcal{L}}}
\newcommand{\hanaB}{\mathcal{B}}
\newcommand{\hanaBhat}{\wh{\mathcal{B}}}
\newcommand{\Psicl}{\Psi_{c\ell}}
\newcommand{\ket}[1]{| #1 \rangle}
\newcommand{\bra}[1]{\langle #1 |}
\newcommand{\verlim}[1]{\varlimsup_{#1}}
\newcommand{\dderiv}[3]{\frac{d^{#3} #1}{d #2^{#3}}}
\newcommand{\fock}{\ket{\Omega}}
\newcommand{\eom}{\Delta_{\text{EOM}}}
\newcommand{\W}{W(\calV,\Psi)}
\newcommand{\wVEV}[1]{\langle\!\langle #1 \rangle\!\rangle}
\newcommand{\Ds}{\Delta_s}
\newcommand{\chL}{\gamma_L}
\newcommand{\chR}{\gamma_R}
\newcommand{\toss}{\calL[G(K)]}
\newcommand{\main}{\calT_{\text{I}}}
\newcommand{\other}{\calT_{\text{I\!I}}}
\newcommand{\Vmid}{\calV_{\text{mid}}}
\newcommand{\KBc}{K\!Bc}
\newcommand{\Vend}{\calV_{\text{end}}}
\begin{titlepage}

\title{
\hfill\parbox{3cm}{\normalsize KUNS-2456}\\[1cm]
{\Large\bf
Inversion Symmetry of Gravitational Coupling \\
in Cubic String Field Theory
}}

\author{
Hiroyuki {\sc Hata}\footnote{
{\tt hata@gauge.scphys.kyoto-u.ac.jp}}
\ and
Toshiko {\sc Kojita}\footnote{
{\tt kojita@gauge.scphys.kyoto-u.ac.jp}}
\\[7mm]
{\it
Department of Physics, Kyoto University, Kyoto 606-8502, Japan
}
}

\date{{\normalsize July 2013}}
\maketitle

\begin{abstract}
\normalsize
It was found that the canonical energy 
of multi-brane solutions in CSFT constructed by the $KBc$ algebra
has a symmetry under the exchange of $K=0$ and $K=\infty$
(inversion symmetry).
On the other hand, the gauge invariant observable (GIO), 
which is regarded as the energy defined by 
the gravitational coupling of open string,
cannot count the energy from $K=\infty$ and therefore 
is not equal to the canonical energy. 
To resolve this discrepancy, we examine the recent argument of 
Baba and Ishibashi which directly relates the two energies.
We find that the gravitational coupling which is equivalent to 
the canonical energy consists of the GIO and another new term,
and the whole has the inversion symmetry. 

\end{abstract}

\thispagestyle{empty}
\end{titlepage}

\tableofcontents
\section{Introduction}
Bosonic open string theory has a tachyonic mode, and therefore 
the perturbative vacuum is unstable. 
People has believed that there should be a stable vacuum with lower energy
\cite{Sen1,Sen2}.
For exploring the stable vacuum we need an off-shell formulation 
for bosonic open string. 
Cubic string field theory (CSFT) \cite{Wit} described by the action
\begin{align}
S=\frac{1}{2}\int \Psi\QB\Psi+\frac{1}{3}\int \Psi*\Psi*\Psi,
\label{eq:CSFT_action}
\end{align}
is such an off-shell formulation.
In fact, the exact classical solution corresponding to the stable vacuum
(tachyon vacuum) was discovered in \cite{S05}. 
It was found that the energy density of the solution is lower than 
that of the perturbative vacuum by $1/(2\pi^2)$, implying that 
there is no D25-brane. Moreover, it was shown also that
there is no physical open string excitations around the solution \cite{ES}. 
Following the success of finding the tachyon vacuum,
general multi-brane solutions have been studied actively
\cite{SE09,MS1,Taka,HK1,MS2,EM,Arro1,MNT,BI,HK,Masuda1,Masuda2,Arro2}.

Then we are interested in the structure of general multi-brane solutions 
including the tachyon vacuum solution in CSFT. 
If we could unveil the universal mathematical structure of 
multi-brane solutions, we can construct various solutions systematically 
without trial and error.
We have carried out studies in this direction in \cite{HK1,HK}
for static and translationally invariant
 pure-gauge type solutions $\Psi=U\QB U^{-1}$.
The energy density $\calE$ of a pure-gauge type solution 
is given as $\calE=\calN/(2\pi^2)$ in terms of $\calN$ defined by
\footnote{
We have put the space time volume equal to one. 
}
\begin{align}
\calN=\frac{\pi^2}{3}\int\left(U\QB U^{-1}\right)^3.
\label{eq:def_calN}
\end{align}
Since the energy density of a single D25 brane is $1/(2\pi^2)$,
$\calN$ of multi-brane solutions should be integers.
We have focused on the fact that the CSFT action \eqref{eq:CSFT_action}
takes the same form as that of the 
Chern-Simons (CS) theory in three dimensions.
Then, $\calN$ corresponds to the winding number 
$N=1/(24\pi^2)\int_{M} \text{tr}\left[(gdg^{-1})^3\right]$ 
in CS theory from the manifold $M$ to the gauge group.
Since the winding number $N$ is quantized to integers,
it is expected that $\calN$ is also quantized 
and has a meaning as a kind of winding number.
In fact, $\calN$ is a topological quantity invariant under an 
infinitesimal deformation
$\delta U=-\lambda U$, which corresponds to 
the gauge transformation
$\delta_{\lambda}\Psi=\QB\lambda+\CR{\Psi}{\lambda}$.
And $\calN$ can be rewritten as the integration of a BRST exact quantity
\cite{HK1},
\begin{align}
\calN=\int \QB\calA\left[U\right].
\label{eq:calN_int_QB_calA}
\end{align}
Therefore, $\calN$ is almost zero and can take a finite value 
due to singularities of $\calA$.  

We evaluated $\calN$ for $U$ written in terms of $K,B,c$
satisfying the ``$\KBc$ algebra'' \cite{Okawa}:
\begin{align}
&\CR{K}{B}=0,\quad \ACR{B}{c}=1,\quad B^2=c^2=0,\nn\\
&\QB K=0,\quad \QB B=K,\quad \QB c=cKc.
\label{eq:KBc_algebra}
\end{align}
Roughly, $K$ and $B$ are line integrals of the energy momentum tensor
and the anti-ghost, respectively, while $c$ is the ghost. 
Concretely, we considered $U$ of the following form 
specified by a function $G(K)$:
\begin{align}
U=1-Bc(1-G(K)),\quad U^{-1}=1+\frac{1}{G(K)}Bc(1-G(K)).
\label{eq:U_general_form}
\end{align}
It is known that the tachyon vacuum solution 
can be written in this form \cite{SE09}.
We found that for a rational function $G(K)$ with the following behaviors
\begin{align}
G(K)\sim 
\begin{cases}
\,\,K^{n_0}&(K\to 0)\\
\,\,(1/K)^{n_\infty}&(K\to \infty)\\
\end{cases},
\label{eq:n0_ninf}
\end{align}
and having no zeros/poles in $\text{Re}K>0$,
$\calN$ does not depend on the details of $G(K)$,
but is determined only by integers $n_0$ and $n_\infty$:
\begin{align}
\calN&=-n_0-n_{\infty}+A(n_0)+A(n_\infty),
\label{eq:calN_n0_ninf}
\end{align}
where $A(n)$ is given by a confluent hypergeometric function
(see eq.(1.12) in \cite{HK}). 
In terms of the expression \eqref{eq:calN_int_QB_calA},
singularities of $\calA[U]$ at $K=0$ and $K=\infty$ determine $\calN$.
$A(n)$ vanishes only for $n=0,\pm 1$ and takes non-integer values for 
other $n$.
This implies that we obtain multi-brane solutions only for 
$\calN=0,\pm 1,\pm 2$ ($\calN=-2$ is the ghost brane solution).\footnote{
In \cite{HK} we proposed a way of constructing solutions 
with $\calN=\pm 3,\pm 4,\cdots$
by using $U$ \eqref{eq:U_general_form} with a rational function $G(K)$.
}
Therefore, in the rest of this paper, we treat only $G(K)$ with  
$n_0,n_\infty=0,\pm 1$ and call the corresponding $\Psi$
the multi-brane solution.

From \eqref{eq:calN_n0_ninf},
we notice that $\calN$ has an invariance under the exchange 
of the origin and the infinity of $K$. 
We call it the inversion symmetry.
This phenomenon is a consequence of a more general property of the 
correlators as we explain below.
Then let us recall that there are two definitions of energy 
in local field theories.  
One is the canonical energy obtained as the Noether charge,
and the other is the one read from the gravitational coupling.  
In CSFT, our $\calN$ corresponds to the former for 
static configurations.\footnote{
Defining the canonical energy in CSFT as the Noether charge of
center-of-mass time translation is a non-trivial problem
since CSFT contains an infinite number of time derivatives in its
interaction term. Here, we are interested only in static solutions, for
which the negative of the action is regarded as the energy.
}
On the other hand,  
what is called the gauge invariant observable (GIO) \cite{HI,GRSZ}
has been proposed
as the energy from the gravitational coupling.
It is natural that the two definitions of energy are equivalent.
In fact,
the equivalence was verified for the tachyon vacuum \cite{Ell,KKT},
and also for multi-brane solutions 
with $n_{\infty}=0$ \cite{MS1,MS2}.\footnote{
Precisely, as seen by comparing \eqref{eq:calN_n0_ninf} and
\eqref{eq:GIO_n0}, the equivalence holds only for 
$n_0=0,\pm 1$ and $n_\infty=0$. In \cite{MS1,MS2} the anomaly term $A(n)$
is missing from $\calN$.    
}
However the inversion symmetry is not realized in the GIO.   
This is because, as will be explained in detail below, 
the GIO cannot detect $n_\infty$, namely, the singularity at $K=\infty$.
It is the purpose of this paper to resolve the discrepancy 
between the canonical energy and the gravitational one for 
solutions with non-trivial $n_\infty$. 

Now, we will explain our problem in more detail. 
First is the inversion transformation. 
It is defined by
\begin{align}
K\to\wt{K}=\frac{1}{K},\qquad
B\to\wt{B}=\frac{B}{K^2},\qquad 
c\to\wt{c}=cK^2 Bc.
\label{eq:inversion_trans}
\end{align}
This transformation exchanges $K=0$ and $K=\infty$ 
by keeping the $\KBc$ algebra \eqref{eq:KBc_algebra}.\footnote{
In general, the $\KBc$ algebra is maintained by the transformation
$\wt{K}=g(K)$  together with 
$\wt{B}=g(K)B/K$ and $\wt{c}=c(K/g(K))Bc$
for an arbitrary $g(K)$ \cite{Erl1,MNT,Erl2}.}
Moreover, we proved in \cite{HK} 
that the $\KBc$ correlators are kept 
invariant under the inversion transformation \eqref{eq:inversion_trans}:
\begin{align}
&\int BcF_1(K)cF_2(K)cF_3(K)cF_4(K)=
\int \wt{B}\wt{c}F_1(\wt{K})\wt{c}F_2(\wt{K})\wt{c}
F_3(\wt{K})\wt{c}F_4(\wt{K}),
\label{eq:marvelous_theorem}
\end{align}
where $F_i(K)$ are arbitrary.
Since the effect of the inversion transformation on $\Psi$
is to replace $G(K)$ with $G(1/K)$ (note that $\wt{B}\wt{c}=Bc$
in \eqref{eq:U_general_form}), $\calN$ is invariant  under $G(K)\to G(1/K)$.
Inversion symmetric expression \eqref{eq:calN_n0_ninf} is a 
consequence of this marvelous property.

By contrast, the property \eqref{eq:marvelous_theorem} 
cannot be applied to the GIO, which is given by 
\begin{align}
\int \calV_{\text{mid}}\Psi\quad\text{with}\quad
\calV_{\text{mid}}=\frac{2}{\pi i}c\p X(i\infty)\bar{c}\bar{\p}X(-i\infty),
\label{eq:def_GIO}
\end{align}
where $\calV_{\text{mid}}$ is the on-shell closed string vertex at the 
string midpoint $(i\infty,-i\infty)$ in the sliver frame.
This GIO represents the interaction where the open string endpoints
join to form a closed string (see fig.\ref{fig:int_calV_Psi} 
in Sec.\,\ref{sec:relation_calN_calV}).
Since the GIO contains explicitly the matter operator $X$, 
it is outside the applicability of the property 
\eqref{eq:marvelous_theorem}.

Evaluation of the GIO for the pure-gauge type solutions 
with a rational function $G(K)$
by using the $\Ke$-regularization (see below) leads to
$n_\infty$ independent result \cite{MS1,MS2}:
\begin{align}
2\pi^2\!\!\int\!\! \calV_{\text{mid}}\Psi[G(K)]
=-\lim_{z\to 0}\frac{z}{G(z)}\p_z G(z)=-n_0.
\label{eq:GIO_n0}
\end{align}
Namely, the GIO does not treat $K=0$ and $K=\infty$ equally. 
Are the canonical energy and the gravitational coupling inequivalent  
to each other for multi-brane solutions?
Or is the GIO insufficient as the gravitational coupling?

A key to a resolution to this problem was given in \cite{BI}.
They gave a direct relation between the canonical energy and the GIO 
for general solutions. 
It is 
\begin{align}
\frac{\calN}{2\pi^2}=\int \calV_{\text{mid}}\Psi+(\text{EOM-terms}),
\label{eq:calN=GIO_EOM}
\end{align}
where the concrete expression of the EOM-terms will be given 
in Sec.\ref{sec:relation_calN_calV}.
When the EOM-terms vanish, this relation implies that $\calN$ 
is equal to the GIO.
We confirm that the EOM-terms for multi-brane solutions vanish
in appendix \ref{sec:EOM_term}.
Therefore, there is contradiction between \eqref{eq:calN=GIO_EOM}
 and the results \eqref{eq:calN_n0_ninf} and \eqref{eq:GIO_n0}.
We examine the proof of \eqref{eq:calN=GIO_EOM} for especially our 
multi-brane solutions and identify a point where the relation 
\eqref{eq:calN=GIO_EOM} breaks down.
As a result, we find that another term needs to be added to the RHS of 
\eqref{eq:calN=GIO_EOM} which detects the singularity at $K=\infty$
(see \eqref{eq:improve_N=GIO}).  
Then the RHS with this improvement is manifestly inversion symmetric.
We claim that the genuine gravitational coupling should be given by 
the GIO plus our newly added term for multi-brane solutions. 

Before finishing the Introduction, we comment on 
the regularization and the EOM.
In order to treat properly a topological quantity $\calN$ as given 
by \eqref{eq:calN_int_QB_calA}, we need a regularization for singularities
at $K=0$ and $\infty$. 
We introduce one regularization common to all the solutions.
For regularizing the $K=0$ singularity, we make the replacement 
\begin{align}
K\to K_{\eps}:= K+\eps,
\label{eq:K_eps_regularization}
\end{align}
in correlators written in terms of $K,B$ and $c$.
We call \eqref{eq:K_eps_regularization}  the $K_{\eps}$-regularization.
Since the eigenvalues of $K$ are non-negative,
$K_{\eps}$-regularization works well.
Next, the regularization for $K=\infty$ is naturally obtained 
as the inversion transformation \eqref{eq:inversion_trans} 
of the $\Ke$-regularization.
Combining the both and introducing the 
regularization parameter $\eta>0$ for $K=\infty$, 
our regularization is finally given by (see \cite{HK} for details)
\begin{align}
K\to K_{\eps\eta}=\frac{K_{\eps}}{1+\eta K_{\eps}},\quad
B\to B_{\eps\eta}=\frac{B}{(1+\eta K_{\eps})^2},\quad
c\to c_{\eps\eta}=c(1+\eta K_{\eps})^2 Bc.
\label{eq:K_eps_eta_regularization}
\end{align}
If the correlator contains $\QB$ we make this replacement 
after evaluating the operation of $\QB$ by using \eqref{eq:KBc_algebra}. 
After the regularization, $\Psi$ is no longer a pure gauge:
\begin{align}
\Psi_{\eps\eta}:=\left[U\QB U^{-1}\right]_{K\to K_{\eps\eta},B\to B_{\eps\eta},
c\to c_{\eps\eta}}
\neq U_{\eps\eta}\QB U^{-1}_{\eps\eta},
\label{eq:Psi_eps_eta}
\end{align}
where $U_{\eps\eta}$ is $U$ 
with the replacement \eqref{eq:K_eps_eta_regularization}.
Then we have to examine whether the regularized $\Psi_{\eps\eta}$ 
satisfies the EOM. 
Namely, it is a non-trivial problem for what kind of test states $\Phi$
the EOM $\int \Phi*\Gamma=0$ with
\begin{align}
\Gamma:=\QB\Psi_{\eps\eta}+\Psi_{\eps\eta}*\Psi_{\eps\eta},
\label{eq:def_Gamma}
\end{align}
holds in the limit $\eps,\eta\to 0$.
In order for $\calN$ to be directly related to energy density $\calE=-S$,
the EOM must hold against $\Phi=\Psi_{\eps\eta}$. 
Therefore, we examined in \cite{HK1,HK} the EOM in the strong sense,
$\text{EOM-test}=\int \Psi_{\eps\eta}*\Gamma$, to find that 
it is given by an inversion symmetric quantity:
\begin{align}
\text{EOM-test}=B(n_0)+B(n_\infty).
\label{eq:EOM_test}
\end{align} 
The function $B(n)$ is equal to zero only when $n=0,\pm 1$ and 
the EOM-test vanishes for the multi-brane solutions 
with $\calN=0,\pm 1 ,\pm 2$. 
The EOM-terms in \eqref{eq:calN=GIO_EOM} consists of the present EOM-test
as well as the EOM against other $\Phi$'s.

The organization of the rest of this paper is as follows.
In Sec.\,\ref{sec:relation_calN_calV}, 
we summarize the derivation of the basic relation 
\eqref{eq:calN=GIO_EOM}
given in \cite{BI}.
In Sec.\,\ref{sec:concrete_eval}, 
which is the main part of this paper, we examine the
derivation of \eqref{eq:calN=GIO_EOM} for the multi-brane solutions. 
There, we find that we have to add a new term to the RHS of 
\eqref{eq:calN=GIO_EOM} which can count $n_\infty$
(see \eqref{eq:improve_N=GIO}).
We summarize the paper in Sec.\,\ref{sec:conclusion}.
In the appendices, various technical details used in the text are given.

\section{Relation between $\calN$ and the GIO 
(Review of \cite{BI})}
\label{sec:relation_calN_calV}

In this section, we briefly review the derivation of the relation 
\eqref{eq:calN=GIO_EOM} between the canonical energy and the gravitational 
coupling given in \cite{BI}. 

First we introduce the notion of string field with width $L$.
Let us take the pure-gauge type solution $\Psi$ with 
$U$ given by \eqref{eq:U_general_form}:
\begin{align}
\Psi=cF(K)BcH(K),
\label{eq:general_Psi}
\end{align}
with
\begin{align}
F(K)=\frac{K}{G(K)},\quad H(K)=1-G(K).
\label{eq:def_FH}
\end{align}
In this paper, we adopt for computational easiness
non-hermitian $\Psi$ \eqref{eq:general_Psi}, which is related to
hermitian $\Psi=\sqrt{H}cFBc\sqrt{H}$ used in \cite{BI} by a gauge
transformation.
Expressing $F$ and $H$ in terms of their inverse Laplace transforms
$f(L)$ and $h(L)$ as 
\begin{align}
F(K)=\int_0^\infty \!\! dL \, e^{-LK}f(L),\quad
H(K)=\int_0^\infty \!\! dL \, e^{-LK}h(L),
\end{align}
the string field $\Psi$ is given as an integration with respect to $L$:
\begin{align}
\Psi=\int_0^\infty \!\! dL\, \wt{\Psi}(L),
\end{align}
where $\wt{\Psi}(L)$ is defined by
\begin{align}
\wt{\Psi}(L)=\int_0^L dL' \, f(L')h(L-L')\,c\,e^{-L'K}Bc\,e^{-(L-L')K}.
\end{align}
We call $\wt{\Psi}(L)$ the string field with width $L$ 
since it represents a strip of width $L$ with ghost insertions.
Quite similarly, we express any string field $\Phi$
consisting of $K,B,c$ and carrying any ghost number as the integration
over the width $L$;
$\Phi=\int_0^\infty\!\! dL \, \wt{\Phi}(L)$.

\begin{figure}[htbp]
\centering
\epsfxsize=0.33\textwidth
\epsfbox{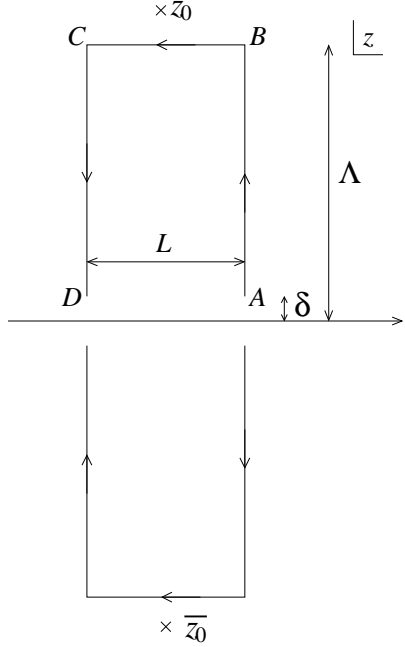}
\caption{
The integration path $P_{L,\Lambda,\delta}$ (in $\text{Im}\,\,z>0$)
and its conjugate $\bar{P}_{L,\Lambda,\delta}$ for $\calG$ \eqref{eq:calG}
acting on the string field with width $L$.
The paths AB and CD correspond to the right and the left half of the string,
respectively. The horizontal path BC goes to the midpoint 
at $z=i \infty$ in the limit $\Lambda\to \infty$. 
}
\label{fig:path_P}
\end{figure}

The key equation for the relation between  $\calN$ and the GIO is 
the dilatation property of the two and three string vertices:  
\begin{align}
&\int (\calG\Psi_1)*\Psi_2+\int \Psi_1*(\calG\Psi_2)=\int\Psi_1*\Psi_2,
\label{eq:calGV2=V2}\\
&\int \left(\calG\Psi_1\right)*\Psi_2*\Psi_3+
\int \Psi_1*\left(\calG\Psi_2\right)*\Psi_3+
\int \Psi_1*\Psi_2*\left(\calG\Psi_3\right)
=\int \Psi_1*\Psi_2*\Psi_3,
\label{eq:calGV3=V3}
\end{align}
where $\calG$ is the dilatation operator of the 
time component $X^0$, which 
we write as $X$ for simplicity.\footnote{
The present $\calG$ is twice the usual dilatation operator.
It satisfies
$\CR{\calG}{X}=-2X,\ \CR{\calG}{P}=2P$.
}
On the sliver frame, the first term on the LHS of \eqref{eq:calGV3=V3},
for example, is given by
\begin{align}
\int\left(\calG\Psi_1\right)*\Psi_2*\Psi_3=
\int_0^\infty\!\!\!\!dL_1\int_0^\infty\!\!\! dL_2\int_0^\infty\!\!\!\!dL_3
\VEV{\bigl(\calG\wt{\Psi}_1(L_1)\bigr)\wt{\Psi}_2(L_2)
\wt{\Psi}_3(L_3)}_{L_1+L_2+L_3},
\label{eq:calGPsi_L}
\end{align}
where $\VEV{\cdots }_s$ denotes the correlator on the infinite cylinder 
of width $s$ (correlators in the sliver frame are summarized 
in appendix\,\ref{sec:correlator}).
On the RHS of \eqref{eq:calGPsi_L}, 
$\calG$ acting on a string field of width $L$ is given by
\begin{align}
\calG=\int_{P_{L,\Lambda,\delta}}\frac{dz}{2\pi i}g_z(z,\bar{z})
-\int_{\bar{P}_{L,\Lambda,\delta}}\frac{d\bar{z}}{2\pi i}g_{\bar{z}}(z,\bar{z}),
\label{eq:calG}
\end{align}
with
\begin{align}
g_z(z,\bar{z})&=2:\left(X(z,\bar{z})-X(z_0,\bar{z}_0)\right)\p X(z):,\nn\\
g_{\bar{z}}(z,\bar{z})&=2:\left(X(z,\bar{z})-X(z_0,\bar{z}_0)\right)
\bar{\p} X(\bar{z}):.
\label{eq:def_g}
\end{align}
The normal ordering $:\quad :$ in \eqref{eq:def_g} 
removes the divergence in
$X\p X$ (and $X\bar{\p}X$) at the coincident point, 
while it has no effect on $X(z_0,\bar{z}_0)\p X(z)$ 
and $X(z_0,\bar{z}_0)\bar{\p} X(\bar{z})$:
\begin{align}
:X(z,\bar{z})\p X(z):&=\lim_{\eps\to 0} \left[X(z,\bar{z})\p X(z+\eps)
-\frac{1}{2\eps}\right],\nn\\
:X(z,\bar{z})\bar{\p} X(\bar{z}):
&= \lim_{\eps\to 0} \left[X(z,\bar{z})
\bar{\p} X(\bar{z}+\eps)
-\frac{1}{2 \eps}\right].
\label{eq:normalordering}
\end{align}
The paths of integration in \eqref{eq:calG}, 
$P_{L,\Lambda,\delta}$ and $\bar{P}_{L,\Lambda,\delta}$,
are given in fig.\ref{fig:path_P}.
Originally, $\calG$ should be defined by an integration along an open string;
namely, from one endpoint (on the real axis in the sliver frame)  
through the midpoint (at $z=\pm i\infty$), to another endpoint.
However the paths for \eqref{eq:calG} are deformed to avoid the 
midpoint and the endpoints by parameters $\Lambda$ and $\delta$, respectively
(see fig.\ref{fig:path_P}).
For $\delta>0$, $z$ and $\bar{z}$ never coincide each other, 
and therefore the normal ordering in \eqref{eq:normalordering} is 
sufficient for making  $g_z$ and $g_{\bar{z}}$ finite.
In the end of calculation, we take the limits 
$\Lambda\to \infty$ and $\delta\to 0$.
The $X(z_0,\bar{z}_0)$ term in \eqref{eq:def_g} is necessary for 
the validity of the BRST Ward-Takahashi identity of the correlators 
(see appendix \ref{sec:BRST_W_T_identity}).  
In order for \eqref{eq:calGV2=V2} and \eqref{eq:calGV3=V3} to hold,
$z_0$ must be driven away to the midpoint $i\infty$ in such a way that 
$\Lambda_0:=\text{Im}z_0$ satisfies $\Lambda_0-\Lambda\to \infty$.

Let us sketch the proof of \eqref{eq:calGV2=V2} and 
\eqref{eq:calGV3=V3} given in \cite{BI}.
First, the contribution from the vertical part of the integration paths
cancel among the terms on the LHS. (This is the case if $z_0$ is common to 
all $\Psi_i$ $(i=1,2,3)$. If we take a different $z_0$ for each $\Psi_i$, 
the cancellation occurs by taking $z_0$'s to the midpoint.) 
Since $\Psi_i$ does not contain $X$ explicitly and $\calG$ does not 
contain the ghosts,
the correlator \eqref{eq:calGPsi_L} factorizes into the 
product of the $X$ part
$\VEV{\calG_{L_i}}_{s}$ and the ghost part
$\langle \wt{\Psi}_1\wt{\Psi}_2\wt{\Psi}_3\rangle_{s}$ with $s=L_1+L_2+L_3$.
The former is given explicitly by  
\begin{align}
\VEV{\calG_{L_i}}_s&=\frac{L_i}{s}
\coth \frac{2\pi \Lambda}{s}
+O\left(e^{-2\pi \Lambda_0/s}\right),
\label{eq:VEV_calG}
\end{align}
where we have used \eqref{eq:VEV_g_z_g_bz}
and that $\Lambda_0-\Lambda \to \infty$.
Summing \eqref{eq:VEV_calG} over $i=1,2,3$ we obtain 
$((L_1+L_2+L_3)/s)\coth 2\pi\Lambda/s\to 1$ as $\Lambda\to\infty$
and therefore \eqref{eq:calGV2=V2} and 
\eqref{eq:calGV3=V3} hold.\footnote{
Here we have taken the limit $\Lambda\to \infty$ before the $L_i$ 
integrations in \eqref{eq:calGPsi_L}.
If the integration over $s=\sum L_i$ has a non-trivial contribution 
from the region $s>\Lambda$, we have to take care of the order 
of the $s$-integration and the limit $\Lambda\to \infty$. 
Fortunately, we do not need to worry about this 
for the multi-brane solutions.  
This is because the integration 
in the region $s>\Lambda$ vanishes in $\Lambda\to \infty$
since $\langle\wt{\Psi}\wt{\Psi}\wt{\Psi}\rangle_s$ decays exponentially
for a large $s$.
}

Then, $\Vmid$ in the GIO \eqref{eq:def_GIO} arises from $\calG$ as follows.
First, the BRST transform of $\calG$ is given by
\begin{align}
\CR{\QB}{\calG}=\chi_L-\chi_R,
\label{eq:QBcalG=chi}
\end{align}
where $\chi_L$ ($\chi_R$) is from the left (right) 
half of the string.\footnote{
$\chi_L$ ($\chi_R$) corresponds to $\chi$ ($\chi^{\dagger}$) 
in \cite{BI}.}
A further action of $\QB$ on $\chi_L$ and $\chi_R$ produces a midpoint 
operator, which is nothing but $\Vmid$ of \eqref{eq:def_GIO}:
\begin{align}
\ACR{\QB}{\chi_L}=-\calV_{\text{mid}},\qquad 
\ACR{\QB}{\chi_R}=-\calV_{\text{mid}}.
\label{eq:QBchi=V}
\end{align}
Recall that the integration path $P_{L,\Lambda,\delta}$ for the 
regularized $\calG$ consists of three parts: 
two vertical parts and the horizontal one on $\text{Im}z=\Lambda\gg 1$
(see fig.\,\ref{fig:path_P}).
Precisely, $\chi_L$ is defined as the sum of 
an integration along the left vertical path DC and the operator 
at the left endpoint D \cite{BI}. 
$\chi_R$ is exactly the same as $\chi_L$ except that its  
vertical path is horizontally shifted.   
Namely, the contribution from the horizontal path BC is discarded in 
\eqref{eq:QBcalG=chi}, 
which needs to be justified for a given $\Psi$. 
As we will see below, the absence the contribution from the horizontal path 
to \eqref{eq:QBcalG=chi} is an important point in relating $\calN$ to
the GIO.

Now we are ready to relate $\calN$ with the GIO.
The principal strategy is as follows. 
The dilatation property \eqref{eq:calGV3=V3} allows us to 
introduce $\calG$ into $\calN$, and we make $\QB$ appear there
by using the EOM. Then, the BRST properties 
\eqref{eq:QBcalG=chi} and \eqref{eq:QBchi=V}
lead to the expression \eqref{eq:def_GIO}.
Concretely, we start with 
\begin{align}
\frac{1}{3}\int\Psi^3&=\int \Psi^2*\calG\Psi=
\int\Gamma*\calG\Psi-\int(\QB\Psi)*\calG\Psi\nn\\
&=\int\Gamma*\calG\Psi-\int \Psi*\CR{\QB}{\calG}\Psi
-\int\Psi*\calG\QB\Psi,
\label{eq:proof_one}
\end{align}
where $\Psi$ denotes the regularized $\Psi_{\eps\eta}$ \eqref{eq:Psi_eps_eta}
and $\Gamma$ is given by \eqref{eq:def_Gamma}.
In \eqref{eq:proof_one},
we have used \eqref{eq:calGV3=V3} and discarded the ``surface term''
$\int \QB \left(\Psi*\calG\Psi\right)$.\footnote{This is allowed since the 
present $\Psi$ is the regularized one.}
Then, rewriting $\QB\Psi$ in the last term of \eqref{eq:proof_one}
into $\Gamma-\Psi^2$ and using 
\begin{align}
\int\Psi*\calG \Psi^2=\int \Psi^3-\int (\calG\Psi)*\Psi^2
=\frac{2}{3}\int \Psi^3,
\label{eq:proof_two}
\end{align}
obtained from \eqref{eq:calGV2=V2} and \eqref{eq:calGV3=V3},
we find that $\calN$ \eqref{eq:def_calN} with regularization is
expressed as 
\begin{align}
\frac{\calN}{2\pi^2}&=\frac{1}{6}\int\!\Psi^3
=\frac{1}{2}\int\Psi*\CR{\QB}{\calG}\Psi
+\frac{1}{2}\int\Psi*\Gamma-\int(\calG\Psi)*\Gamma.
\label{eq:proof_three}
\end{align}
Using \eqref{eq:QBcalG=chi} and \eqref{eq:QBchi=V}, 
the first term on the RHS of \eqref{eq:proof_three}
is further rewritten as 
\begin{align}
\frac{1}{2}\int \Psi *\CR{\QB}{\calG}\Psi&=
\frac{1}{2}\int\Psi*(\chi_L-\chi_R)\Psi
=\!\int\!\! \chi_L\Psi^2
=\!\int\!\! \chi_L\Gamma-\!\int\!\! \chi_L\QB \Psi\nn\\
&=\!\int \chi_L\Gamma+\!\int\!\! \Vmid\Psi,
\label{eq:proof_four}
\end{align}
where we have used at the second equality 
that $\chi_L$ and $\chi_R$ are Grassmann-odd quantities
defined by the integration of a common quantity
along different vertical paths.
From \eqref{eq:proof_three} and \eqref{eq:proof_four},
we finally obtain
\begin{align}
\frac{\calN}{2\pi^2}=\int\! \Vmid \Psi 
+\frac{1}{2}\int \Psi*\Gamma
-\int\!\!(\calG\Psi)*\Gamma
+\int\!\! \chi_L\Gamma.
\label{eq:proof_five}
\end{align}
This is the precise expression of \eqref{eq:calN=GIO_EOM}.

\begin{figure}
\begin{minipage}[t]{0.47\textwidth}
\centering
\epsfxsize=0.8\textwidth
\epsfbox{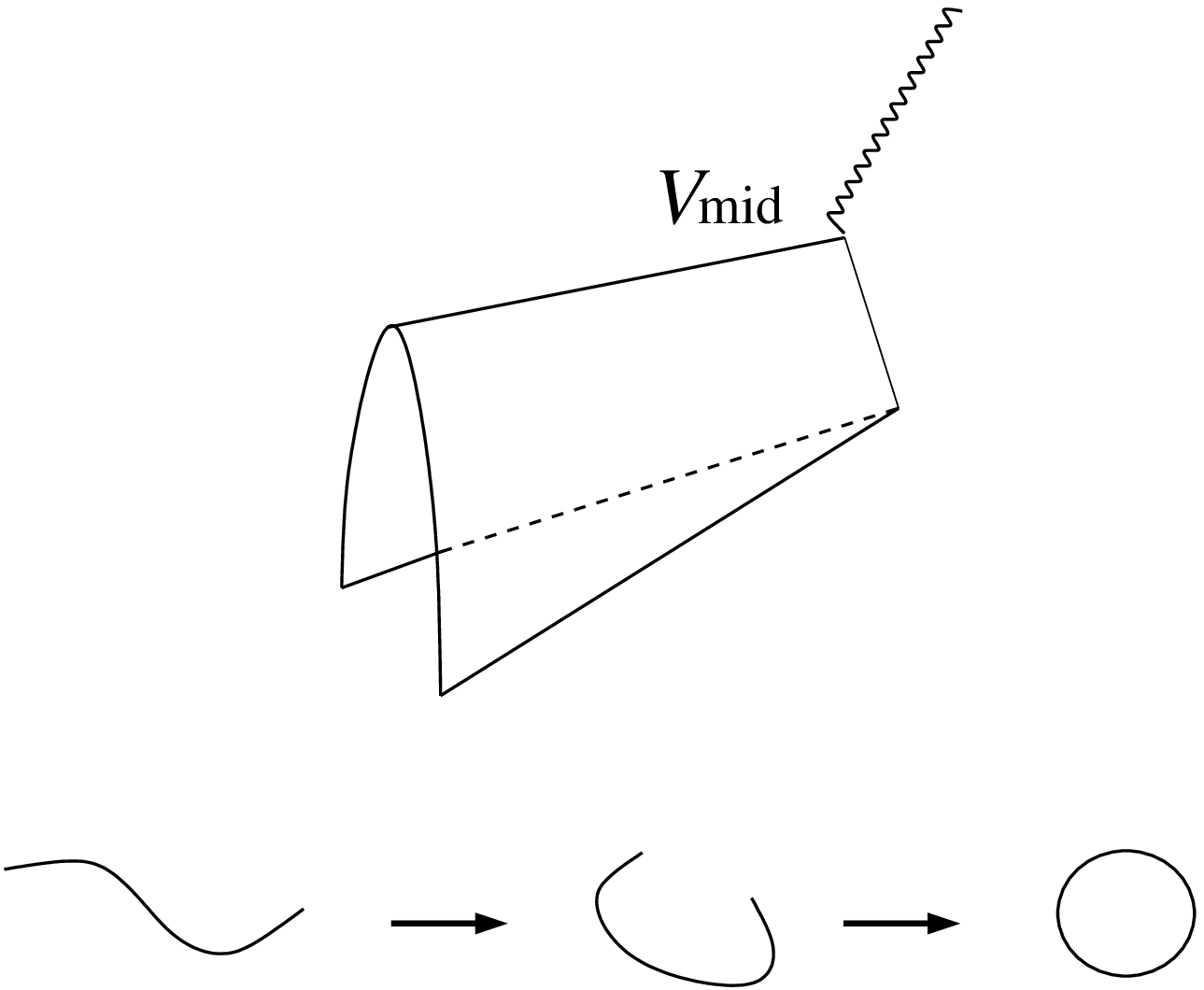}
\caption{The gravitational coupling of open string 
corresponding to the GIO $\int \Vmid\Psi$.
The closed string is coupled to the midpoint of the glued open string
(upper figure).
The lower figure shows an intuitive time development of the process.}
\label{fig:int_calV_Psi}
\end{minipage}
\hfill
\begin{minipage}[t]{0.47\textwidth}
\centering
\epsfxsize=\textwidth
\epsfbox{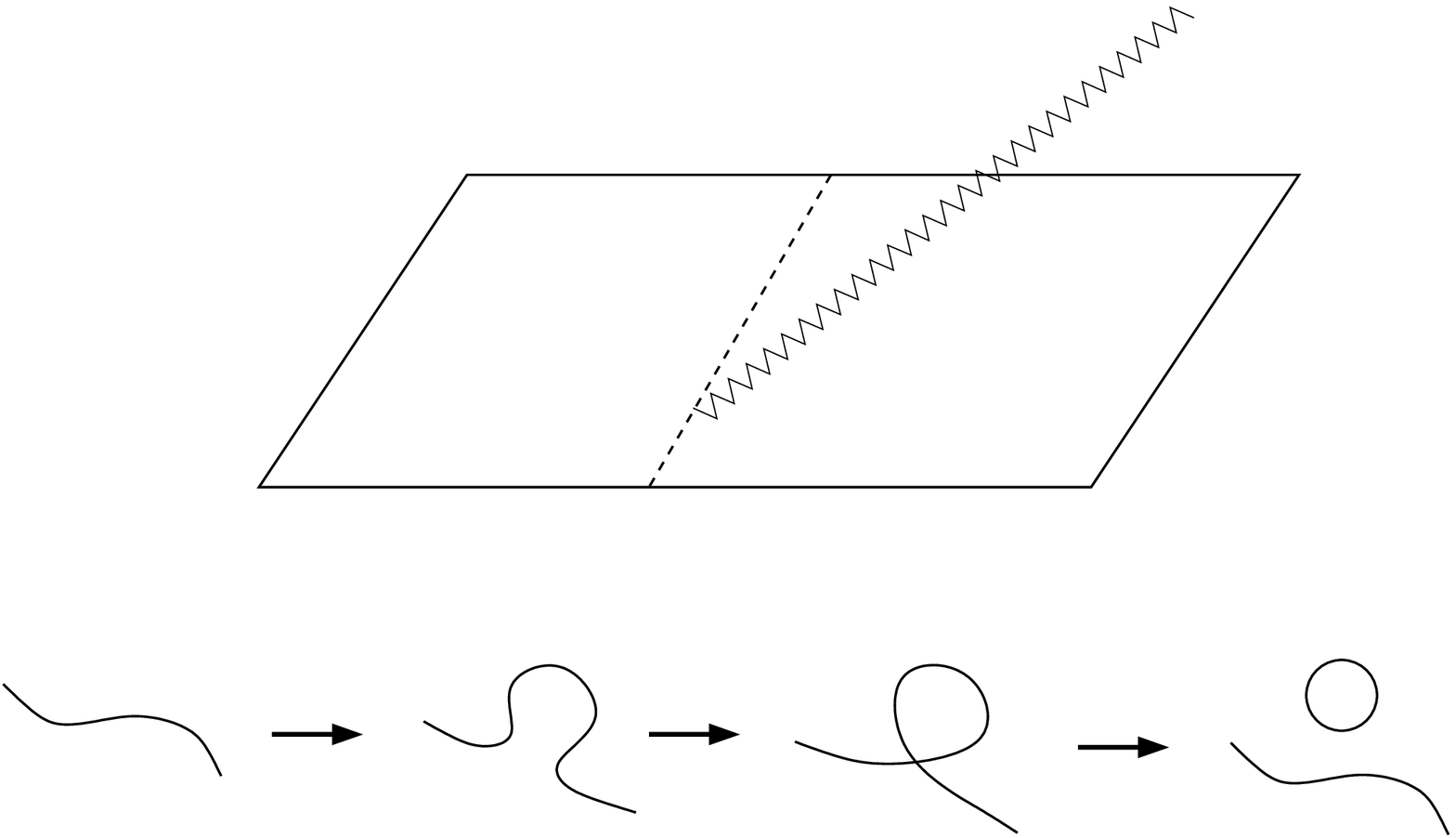}
\caption{The gravitational coupling of open string corresponding to 
$\int \Psi\CR{\QB}{\calG}\Psi$.
Since $\CR{\QB}{\calG}$ is an integrated quantity, 
a closed string is coupled at any point on the open string (upper figure).
The lower figure shows an intuitive time development of the process.}
\label{fig:int_Psi_CR_QB_calG_Psi}
\end{minipage}
\end{figure}

Eq.\,\eqref{eq:proof_five} relates the canonical energy $\calN$ 
with the GIO  \eqref{eq:def_GIO}, namely, the gravitational coupling
of fig.\,\ref{fig:int_calV_Psi}
representing the process an open string converting to a closed string. 
Note that eq.\,\eqref{eq:proof_three} also relates 
$\calN$ with another kind of
gravitational coupling $1/2\int\!\Psi*\CR{\QB}{\calG}\Psi$,\footnote{
It is interesting if we can identify this term
$1/2\int\!\Psi*\CR{\QB}{\calG}\Psi$ as the $00$-component of the
energy-momentum tensor obtained by the variation of the CSFT action
with respect to the background space-time metric.
For this, the following observation might be useful.
Note that, for the BRST operator under the background $g_{\mu\nu}$,
$$
\QB^{(g)}\sim \int d\sigma\bigl\{c(
g^{\mu\nu}P_\mu P_\nu+g_{\mu\nu}(d X^\mu/d\sigma)(d
X^\nu/d\sigma))+\cdots\bigr\},
$$
its variation with respect to $g_{00}$ around the flat metric gives
$1/4\CR{\QB}{\calG}$. Therefore, the variation of the CSFT action
gives the present coupling (up to the multiplying
factor) under the assumption that the star product $*$ and the
integration $\int$ are background independent.
}
which describes the process of fig.\,\ref{fig:int_Psi_CR_QB_calG_Psi}; 
an open string emitting a closed string.

\section{Inversion symmetry of gravitational coupling}
\label{sec:concrete_eval}
Recall that $\calN$ has the inversion symmetry, namely, it is invariant
under the exchange of $K=0$ and $K=\infty$. Concretely, 
for $G(K)$ with the behaviors of \eqref{eq:n0_ninf},
$\calN$ is given by \eqref{eq:calN_n0_ninf} which is 
symmetric under $n_0\leftrightarrow n_\infty$.
The origin of the inversion symmetry is the invariance 
\eqref{eq:marvelous_theorem} of the $\KBc$ correlators under 
the inversion transformation \eqref{eq:inversion_trans}. 
On the other hand, the GIO \eqref{eq:def_GIO},
which is  not expressed by $\KBc$ alone and therefore 
the invariance \eqref{eq:marvelous_theorem}
cannot be applied, does not have the inversion symmetry.
In fact, the value of the GIO is given by \eqref{eq:GIO_n0} 
for $G(K)$ with \eqref{eq:n0_ninf}.
It cannot detect the singularity at $K=\infty$.

Then, \eqref{eq:proof_five},
which claims the equivalence of $\calN$ and the GIO,
is a contradiction 
(the EOM terms containing $\Gamma$ vanish for the multi-brane solutions).
In this section, we examine the process of getting the relation 
\eqref{eq:proof_five}
for our multi-brane solutions.
We identify at which step from \eqref{eq:proof_one} to \eqref{eq:proof_five}
the inversion symmetry fails.
We further obtain the correct expression of the gravitational coupling 
which agrees with $\calN$ for all multi-brane solutions and therefore 
has the inversion symmetry.

Dropping all the EOM terms, the results 
\eqref{eq:proof_one} -- \eqref{eq:proof_five}
are summarized by 
\begin{align}
\frac{\calN}{2\pi^2}&=\frac{1}{2}\int \Psi*\CR{\QB}{\calG}\Psi=
\int \Vmid \Psi.
\label{eq:target}
\end{align}
The first equality is valid since it is a consequence of the dilatation 
property and the BRST Ward-Takahashi identity. 
In the rest of this section, 
we focus on the second equality in \eqref{eq:target}.

\subsection{$\int \Psi *\CR{\QB}{\calG}\Psi$ }
In this subsection, we concretely evaluate 
\begin{align}
\int\Psi*\CR{\QB}{\calG}\Psi,
\label{eq:Psi_QBcalG_Psi}
\end{align}
for the regularized multi-brane solutions $\Psi_{\eps\eta}$.
We omit the subscript $\eps\eta$ in \eqref{eq:Psi_QBcalG_Psi} and 
hereafter.
Let us first consider $\CR{\QB}{\calG}$. 
Its integrand is given, by taking into account the regularization 
for the normal ordering \eqref{eq:normalordering}, by\footnote
{
Corresponding to the $\KBc$ algebra \eqref{eq:KBc_algebra}, we have
$$
\QB X(z,\bar{z})=-\left(c\p +\bar{c}\bar{\p}\right)\!
X(z,\bar{z}),\quad
\QB \p X(z)=-\p\left(c\p X(z)\right) ,\quad
\QB c=- c\p c.
$$
}
\begin{align}
-\CR{\QB}{g_z(z,\bar{z})}&=
2\bar{c}\bar{\p}X(\bar{z})\p X(z)
-2\left(\left(c\p+\bar{c}\bar{\p}\right)X(z_0,\bar{z}_0)\right)\p X(z)\nn\\
&+ 2\lim_{\eps\to 0}\left\{
c\p X(z)\p X(z+\eps)+
\left(X(z,\bar{z})-X(z_0,\bar{z}_0)\right)
\p\left(c\p X(z+\eps)\right)\right\},
\label{eq:QB_calG_one}
\end{align}
where in the second line we have put $\eps=0$ since 
$\p X$'s there are located at different points.
Expressing the last term of \eqref{eq:QB_calG_one} as 
the sum of a total derivative term and the rest, 
and Laurent-expanding \eqref{eq:QB_calG_one} with respect $\eps$
by using the definition \eqref{eq:normalordering} of the normal ordering,
we get a finite result:
\begin{align}
-\CR{\QB}{g_z(z,\bar{z})}&=
2\bar{c}\bar{\p}X(\bar{z})\p X(z)
-2(\left(c\p+\bar{c}\bar{\p}\right)X(z_0,\bar{z}_0))\p X(z)\nn\\
&\quad+2\lim_{\eps\to 0}\left\{
\p X(z)\p X(z+\eps)\left(c(z)-c(z+\eps)\right)\right.\nn\\
&\left.\quad+\p\left[\left(X(z,\bar{z})-X(z_0,\bar{z}_0)\right)
c\p X(z+\eps)\right]\right\}\nn\\
&=2\bar{c}\bar{\p}X(\bar{z})\p X(z)
-2(\left(c\p+\bar{c}\bar{\p}\right)X(z_0,\bar{z}_0))\p X(z)\nn\\
&\quad+ 
\frac{1}{2}\p^2 c(z)
+2\p\left[:\!\left(X(z,\bar{z})-X(z_0,\bar{z}_0)\right)
c\p X(z)\!:\right].
\end{align}
Making a similar manipulation for $\CR{\QB}{g_{\bar{z}}}$, 
we get finally 
\begin{align}
\CR{\QB}{\calG}&=\calC_{\text{I}}+\calC_{\text{IIA}}+\calC_{\text{IIB}},
\label{eq:CR_QB_calG}
\end{align}
where
\begin{align}
\calC_{\text{I}}&=-\int_{P_{L,\Lambda,\delta}}\!\!\frac{dz}{2\pi i}
4\bar{c}\bar{\p}X(\bar{z})\p X(z)
+\int_{\bar{P}_{L,\Lambda,\delta}}\!\!\frac{d\bar{z}}{2\pi i}
4c\p X(z)\bar{\p}X(\bar{z}),
\label{eq:calC_I}\\
\calC_{\text{IIA}}&=2(c\p +\bar{c}\bar{\p})X(z_0,\bar{z}_0)\left(
\int_{P_{L,\Lambda,\delta}}\frac{dz}{2\pi i}\p X(z)
-\int_{\bar{P}_{L,\Lambda,\delta}}
\frac{d\bar{z}}{2\pi i}\bar{\p} X(\bar{z})\right),
\label{eq:calC_IIA}\\
\calC_{\text{IIB}}&=-\frac{1}{2}\int_{P_{L,\Lambda,\delta}}\frac{dz}{2\pi i}\p^2c(z)
+\frac{1}{2}\int_{\bar{P}_{L,\Lambda,\delta}}\frac{d\bar{z}}{2\pi i}\bar{\p}^2
\bar{c}(\bar{z})\nn\\
&\quad-\int_{P_{L,\Lambda,\delta}}\!\! dz\,\p\kappa(z,\bar{z})
-\int_{\bar{P}_{L,\Lambda,\delta}}\!\!
d\bar{z}\,\bar{\p}\kappa(z,\bar{z}),
\label{eq:calC_IIB}
\end{align}
with 
\begin{align}
\kappa(z,\bar{z})=\frac{1}{\pi i}
:\!\left(X(z,\bar{z})-X(z_0,\bar{z}_0)\right)
\left(c\p
  X(z)-\bar{c}\bar{\p}X(\bar{z})\right)\!:.
\label{eq:def_kappa}
\end{align}
Since we have used the same $\kappa(z,\bar{z})$ in
the last two terms of $\calC_{\text{IIB}}$,
the coefficient of $\calC_{\text{I}}$ has been doubled. 
This is the result of \cite{BI}.

We will give some remarks on the evaluation of \eqref{eq:Psi_QBcalG_Psi}.
First, $\calG$ \eqref{eq:calG} and hence $\CR{\QB}{\calG}$ 
\eqref{eq:CR_QB_calG} depends on regularization parameters 
$\Lambda$ and $\delta$.
We have to calculate \eqref{eq:Psi_QBcalG_Psi} 
by keeping them finite and then take the
limits $\Lambda\to\infty$ and $\delta\to 0$ in the end.
As we will see later, the presence of $\delta>0$ is particularly
important for obtaining the correct result.
Next, recall that $\Psi$ is given in the form of \eqref{eq:general_Psi},
and, in particular, it contains one $B$, anti-ghost integration along a 
vertical path. 
Since \eqref{eq:Psi_QBcalG_Psi} contains two $B$,
we reduce it into a $\KBc$ correlator with a single $B$
by using $\ACR{B}{c}=1$ and $B^2=0$.
In removing $B$ in $\Psi$ acted by $\CR{\QB}{\calG}$,
we have to take care of the anti-commutator $\ACR{\CR{\QB}{g_z}}{B}$.
It has a simple expression obtained from the Jacobi identity
and $\CR{g_z}{B}=0$:
\begin{align}
\ACR{\CR{\QB}{g_z}}{B}=\ACR{\CR{g_z}{B}}{\QB}+\CR{\ACR{B}{\QB}}{g_z}=
\CR{K}{g_z}=-\left(\frac{\p}{\p x}+\frac{\p}{\p x_0} \right)g_z, 
\label{eq:Jacobi}
\end{align}
with $z=x+i y$ and $z_0=x_0+i y_0$.
In evaluating the contribution of this anti-commutator term 
to \eqref{eq:Psi_QBcalG_Psi}, we are allowed to take the 
the $X$-expectation value of \eqref{eq:Jacobi}.
Fortunately, this vanishes since the $(x,x_0)$-dependence of 
$\VEV{g_z}_s$ and $\VEV{g_{\bar{z}}}_s$ is only through $x-x_0$
as seen from the formulas
\eqref{eq:VEV_g_z_g_bz}.
Therefore, we may forget $\CR{\QB}{\calG}$ when we reduce 
\eqref{eq:Psi_QBcalG_Psi} into the form with a single $B$.
For $\Psi$ of the form $\Psi=cF(K)BcH(K)$,
we have symbolically  
\begin{align}
\int\!\! \Psi\CR{\QB}{\calG}\Psi=\int\!\! cFBcH\wick{2}{<1{}\, cFBcH\, >1{}\,}
=\int\!\!
cFBcH\wick{2}{<1{}\, cFH\, >1{}\,}
-\int\!\!FBcH\wick{2}{<1{}\, cFcH\, >1{}\,},
\label{eq:wick}
\end{align}
where $\wick{0}{<1{} \qquad  >1{}}$ denotes $\CR{\QB}{\calG}$ and 
its integration path.

In the rest of this subsection,
we calculate \eqref{eq:Psi_QBcalG_Psi} by dividing it into two parts
$\main$ and $\other$ corresponding to 
$\calC_{\text{I}}$ and $\calC_{\text{IIA}}+\calC_{\text{IIB}}$
of \eqref{eq:CR_QB_calG}, respectively.

\subsubsection{Evaluation of $\main$}
\label{sec:eval_T_I}
Here, we evaluate 
\begin{align}
\main&=\int\Psi\left\{-\int_{P_{L,\Lambda,\delta}}\!\!\frac{dz}{2\pi i}
4\bar{c}\bar{\p}X(\bar{z})\p X(z)
+\int_{\bar{P}_{L,\Lambda,\delta}}\!\!\frac{d\bar{z}}{2\pi i}
4c\p X(z)\bar{\p}X(\bar{z})\right\}\Psi,
\label{eq:main_term}
\end{align}
for our multi-brane solutions.
Recall that \eqref{eq:main_term} is given by an integration
like \eqref{eq:calGPsi_L} over the widths $L_i$ of a correlator 
on a sliver frame of width $s=\sum_i L_i$.

First, we see that the correlator  vanishes on the horizontal path
in the limit $\Lambda\to \infty$ for each $s$. 
This is because the correlator is the product of 
$\VEV{\p X(z)\bar{\p}X(\bar{z})}_s\bigl|_{z=x+i\Lambda}$
and a $\KBc$ correlator with one ghost on the horizontal path,
and the former behaves as $e^{-4\pi\Lambda/s}$ for a large $\Lambda$, 
while the $\KBc$ part as $e^{2\pi\Lambda/s}$ as seen from 
\eqref{eq:pXpX}, \eqref{eq:pXbarpX} and \eqref{eq:Bcccc}.\footnote{
There is no subtlety in taking the limit $\Lambda\to \infty$ 
since the integrand of the $s$-integration decays sufficiently
fast at large $s$ for our multi-brane solutions.
}

Therefore, we have only to take care of the $z$ integrations 
along the vertical paths. 
We define the integration $\chL$ along the left 
vertical path from D to C of fig.\,\ref{fig:path_P}:   
\begin{align}
\chL&\equiv \int^{\text{C}}_{\text{D}}\!\!\frac{dz}{2\pi i}
4\bar{c}\bar{\p}X(\bar{z})\p X(z)
-\int^{\bar{\text{C}}}_{\bar{\text{D}}}\!\!\frac{d\bar{z}}{2\pi i}
4c\p X(z)\bar{\p} X(\bar{z}),
\label{eq:gamma_L}
\end{align}
where $\bar{\text{C}}$ and $\bar{\text{D}}$ are the complex conjugate points
of C and D, respectively. 
Defining $\chR$ similarly as an integration from A to B,
we have\footnote{
For \eqref{eq:T=2int_chiPsi}, we should consider the second expression 
of \eqref{eq:wick}; the last expression with a single $B$ is not suited.
} \footnote{
We discuss in Sec.\,\ref{sec:relate_BI} 
the relation between $\chL$ and $\chi_L$ 
of Sec.\,\ref{sec:relation_calN_calV}.
}
\begin{align}
\main=\int\!\!\Psi*\left(\chL-\chR\right)\Psi=
2\int \chL\Psi^2,
\label{eq:T=2int_chiPsi}
\end{align}
where we have used at the second equality 
\begin{align}
\int (\chR \Phi_1)*\Phi_2=(-1)^{|\Phi_1|} \int \Phi_1*\chL \Phi_2,
\end{align}
valid for any string fields $\Phi_{1,2}$ with ghost number $|\Phi_{1,2}|$.
Namely, $\chR \Phi_1$ implies $\Phi_1$ with the insertion of $\chR$ at 
its right side multiplied by the sign factor necessary for exchanging 
$\chR$ and $\Phi_1$. 

\begin{figure}
\centering
\epsfxsize=0.5\textwidth
\epsfbox{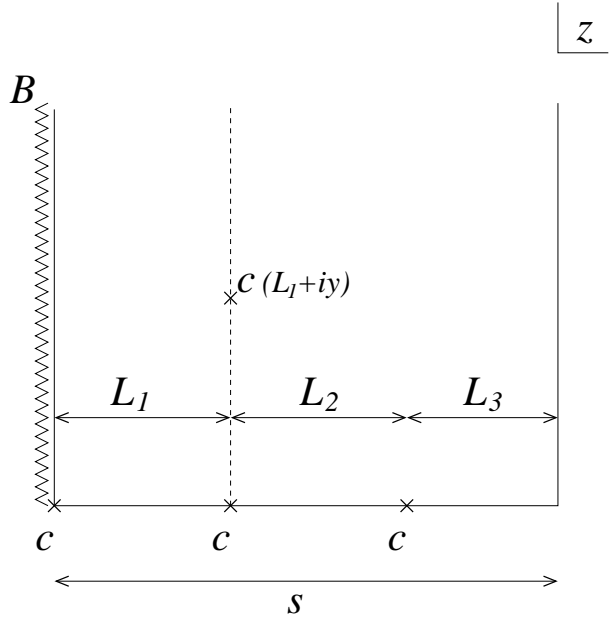}
\caption{Sliver frame for $\int \Psi\CR{\QB}{\calG}\Psi$.
The vertical dashed line corresponds to the vertical path CD in 
fig.\,\ref{fig:path_P}.
}
\label{fig:Psi_QBcalG_Psi}
\end{figure}
Let us evaluate \eqref{eq:T=2int_chiPsi} for the regularized $\Psi_{\eps\eta}$
\eqref{eq:Psi_eps_eta} corresponding to the unregularized $\Psi$ given by
\eqref{eq:general_Psi} with \eqref{eq:def_FH}.
Using the $\KBc$ algebra, $\Psi_{\eps\eta}$ is reduced to the form
$\Psi_{\eps\eta}=cFBcH$
with newly defined $F$ and $H$; 
\begin{align}
F(K)=\frac{K_{\eps}^2}{G(K_{\eps\eta})K_{\eps\eta}},\quad
H(K)=1-G(K_{\eps\eta}),\quad
J(K)=F(1-H)=\frac{K_{\eps}^2}{K_{\eps\eta}}.
\label{eq:def_FHJ}
\end{align}
Here, we have introduced $J$ for later convenience.
Using $\Psi_{\eps\eta}^2=(cFcJ-cJcF)BcH$, 
we obtain (see fig.\,\ref{fig:Psi_QBcalG_Psi})
\begin{align}
\main&=\int\!\!
\left(\prod_{i=1}^3dL_i\right)
h(L_1)\left(f(L_2)j(L_3)-j(L_2)f(L_3)\right)\calW_{\text{I}}(L_1,L_2,L_3),
\label{eq:int_gamma_Psi^2}
\end{align}
with
\begin{align}
\calW_{\text{I}}
&=\frac{8}{\pi}\int^{\Lambda}_{\delta}\!dy\,
\VEV{\p X(L_1+iy)\bar{\p}X(L_1-iy)}
\text{Re}\VEV{Bc(0)c(L_1+iy)c(L_1)c(L_1+L_2)}_s\nn\\
&=\frac{s }{2 \pi ^3}
\left[\tanh \frac{\pi  y}{s}\right]_{y=\delta}^{y=\Lambda} 
\left(L_2 \sin \frac{2 \pi 
   L_1}{s}-L_1 \sin\frac{2 \pi 
   L_2}{s}\right).
\label{eq:def_WI}
\end{align}
Here, $f(L)$, $h(L)$ and $j(L)$ are the inverse Laplace transforms of 
$F(K)$, $H(K)$ and $J(K)$, respectively,
and we have used \eqref{eq:pXpX} and \eqref{eq:real_Bcccc} for 
the correlators.
Using the $sz$-trick \cite{MS1,MS2} for \eqref{eq:int_gamma_Psi^2},
we finally get
\begin{align}
\main&=\int_0^\infty\!\!ds\, \frac{s}{2\pi^3}
\left[\tanh\frac{\pi y}{s}\right]_{y=\delta}^{y=\Lambda}
\int_{-i\infty}^{i\infty}\frac{dz}{2\pi i}e^{sz}\calF(s,z),
\label{eq:TI_term}
\end{align}
where $\calF(s,z)$ is defined by 
\begin{align}  
&\calF(s,z)=
\int\!\!\left(\prod_{i=1}^3dL_i\right)
h(L_1)\left(f(L_2)j(L_3)-j(L_2)f(L_3)\right)
 e^{-z(L_1+L_2+L_3)}\nn\\
&\qquad \qquad\times \left(L_2 \sin \frac{2 \pi 
   L_1}{s}-L_1 \sin\frac{2 \pi 
   L_2}{s}\right)\nn\\
&=\frac{1}{2i}\left\{
(\Delta_s H)\Bigl[J'(z)F(z)-F'(z)J(z)\Bigl]
+H'(z)\Bigl[(\Delta_s F)(z)J(z)-(\Delta_sJ)(z)F(z)\Bigr]
\right\},
\label{eq:def_calF}
\end{align}
with
\begin{align}
(\Delta_s F)(z):=F\left(z-\frac{2\pi i}{s}\right)
-F\left(z+\frac{2\pi i}{s}\right).
\label{eq:def_Delta_s}
\end{align}

\subsubsection{Evaluation of $\other$}
\label{sec:Evalu_T_II}
Next we consider $\other$.
This consists of $\calT_{\text{IIA}}$ and $\calT_{\text{IIB}}$
from $\calC_{\text{IIA}}$ and $\calC_{\text{IIB}}$, respectively.
We start with the former:
\begin{align}
\calT_{\text{IIA}}&=\int\Psi\left\{2(c\p +\bar{c}\bar{\p})X(z_0,\bar{z}_0)\left(
\int_{P_{L,\Lambda,\delta}}\frac{dz}{2\pi i}\p X(z)
-\int_{\bar{P}_{L,\Lambda,\delta}}
\frac{d\bar{z}}{2\pi i}\bar{\p} X(\bar{z})\right)\right\}\Psi.
\label{eq:z0_term}
\end{align}
This is given by \eqref{eq:int_gamma_Psi^2} with $\calW_{\text{I}}$
replaced with
\begin{align}
\calW_{\text{IIA}}&=2
\VEV{Bc(0)c(z_0)c(L_1)c(L_1+L_2)}_s
\int_{P-\bar{P}}\frac{dz}{2\pi i}\VEV{\p X(z_0) \p X(z)}_s+
\left(z_0 \to \bar{z}_0 \right),
\label{eq:def_calWIIA}
\end{align}
where $P-\bar{P}$ is short for $P_{L,\Lambda,\delta}-\bar{P}_{L,\Lambda,\delta}$.
If we put $\delta =0$ from the start,
the path $P-\bar{P}$ is 
reduced to the closed rectangle contour 
BC$\bar{\text{C}}\bar{\text{B}}$.
Since $z_0$ and $\bar{z}_0$ are outside the contour and hence 
the integrand is regular inside it,
\eqref{eq:def_calWIIA} vanishes.
Of course, this is not a correct prescription as we mentioned before.
We have to evaluate (3.23) by keeping $\delta$ finite.
For a finite $\delta$ and $z_0=x_0+i\Lambda_0$,
we have
\begin{align}
&\int_{P-\bar{P}}dz\,\VEV{\p X(z_0) \p X(z)}_s\nn\\
&=\frac{\pi}{2s}\left(
\cot\frac{\pi (x_0-L_1+i(\Lambda_0-\delta))}{s}
-\cot \frac{\pi (x_0-L_1+i(\Lambda_0+\delta))}{s}\right)\nn\\
&\quad+\left(\text{terms from }\text{A and }\bar{\text{A}}\right),
\label{eq:z0_term_1}
\end{align}
where we have written explicitly the contribution from the points D and 
$\bar{\text{D}}$. (Note that the point D corresponds to $L_1+i\delta$. 
See figs.\,\ref{fig:path_P} and \ref{fig:Psi_QBcalG_Psi}.)
For a large $\Lambda_0$, \eqref{eq:z0_term_1} decreases like 
$e^{-2\pi\Lambda_0/s}$.
On the other hand, the $\KBc$ correlator in \eqref{eq:def_calWIIA} 
blows up like $e^{2\pi\Lambda_0/s}$.
As a result, \eqref{eq:def_calWIIA} is finite in the limit 
$\Lambda_0\to \infty$:
\begin{align}
\calW_{\text{IIA}}&=\frac{s}{\pi^3}
\sinh \frac{2\pi \delta}{s}\left(L_2 \sin \frac{2
   \pi  L_1}{s}-L_1 \sin \frac{2 \pi 
   L_2}{s}\right)\qquad (\Lambda_0\to \infty),
\label{eq:result_calWIIA}
\end{align}
where we have used that the contribution of the terms from 
A and $\bar{\text{A}}$ in \eqref{eq:z0_term_1} 
is the same as that of the D and $\bar{\text{D}}$ term. 
This $\calW_{\text{IIA}}$ indeed vanishes if we put $\delta=0$.
However, the limit $\delta\to 0$ must be taken in the end, and therefore 
$\sinh(2\pi\delta/s)$ in \eqref{eq:result_calWIIA} makes the
$s$-integration divergent at $s=0$ for our multi-brane solutions.

Finally, we consider $\calT_{\text{IIB}}$. 
Since $\calC_{\text{IIB}}$ is given as a
surface term, we have 
\begin{align}
\calT_{\text{IIB}}&=-\int\Psi \left\{\left[\frac{1}{4\pi i}
\left(\p c(z)-\bar{\p}c(\bar{z})\right)
+\kappa(z,\bar{z})\right]_{(z,\bar{z})=(A,\bar{\text{A}})}
^{(z,\bar{z})=(D,\bar{\text{D}})}\right\}\Psi.
\label{eq:surface_term}
\end{align}
Using that the (A,$\bar{\text{A}}$) term contributes the same as 
the (D,$\bar{\text{D}}$) term,
$\calT_{\text{IIB}}$ is given by \eqref{eq:int_gamma_Psi^2} with 
$\calW_{\text{I}}$ replaced by
\begin{align}
\calW_{\text{IIB}}&=
\left(\frac{1}{2\pi }\deriv{}{y}
+\frac{1}{s}\coth\frac{2\pi y}{s}\right)
2\,\text{Re}\VEV{Bc(0)c(L_1+iy)c(L_1)c(L_1+L_2)}_s\bigl|_{y=\delta}.
\label{eq:def_WIIB}
\end{align}
The contribution of the $X(z_0,\bar{z}_0)$ term in 
$\kappa$ \eqref{eq:def_kappa}
to $\calW_{\text{IIB}}$ vanishes in the limit $\Lambda_0 \to \infty$
because the dumping factor $e^{-2\pi\Lambda_0/s}$ 
from the $X$-correlator cannot 
be canceled by the $\KBc$ correlator in this case.
Then, using the relation
\begin{align}
\left(\frac{1}{2\pi}\deriv{}{y}-\frac{1}{s}\coth \frac{\pi y}{s}\right)
2\,\text{Re}\VEV{Bc(0)c(L_1+iy)c(L_1)c(L_1+L_2)}_s=0,
\label{eq:deriv_coth}
\end{align}
we obtain
\begin{align}
\calW_{\text{IIB}}
&=-\frac{s}{2\pi^3}\left(
\sinh \frac{2\pi\delta}{s}+\cosh\frac{2\pi\delta}{s}\tanh\frac{\pi\delta}{s}
\right)
\left(
L_2\sin\frac{2\pi L_1}{s}-L_1\sin\frac{2\pi L_2}{s}
\right).
\label{eq:result_calWIIB}
\end{align}
Similarly to \eqref{eq:result_calWIIA},
\eqref{eq:result_calWIIB} makes the $s$-integration for 
\eqref{eq:surface_term} divergent at $s=0$ for a finite $\delta$.
However, we find that the sum of $\calW_{\text{IIA}} $ \eqref{eq:result_calWIIA}
and $\calW_{\text{IIB}} $ \eqref{eq:result_calWIIB}
is a safe function at $s=0$;
\begin{align}
\calW_{\text{IIA}}+\calW_{\text{IIB}}
&=\frac{s}{2\pi^3}\tanh\frac{\pi\delta}{s}
\left(
L_2\sin\frac{2\pi L_1}{s}-L_1\sin\frac{2\pi L_2}{s}\right).
\end{align}
Therefore, $\calT_{\text{II}}=\calT_{\text{IIA}}+\calT_{\text{IIB}}$
is finally given by
\begin{align}
\calT_{\text{II}}&=\int_0^\infty\!\! ds \frac{s}{2\pi^3}
\tanh\frac{\pi \delta}{s}\int_{-i\infty}^{i\infty}\!\frac{dz}{2\pi i}
e^{sz}\calF(s,z).
\label{eq:result_TII}
\end{align}
with $\calF(s,z)$ given by \eqref{eq:def_calF}.

\subsubsection{The total  of $\int \Psi*\CR{\QB}{\calG}\Psi$}
The total of \eqref{eq:Psi_QBcalG_Psi} is now obtained as 
the sum of $\main$ \eqref{eq:TI_term} and $\other$ \eqref{eq:result_TII}:
\begin{align}
\int\Psi*\CR{\QB}{\calG}\Psi&=\calT_{\text{I}}+\calT_{\text{II}}
=\int_{0}^\infty \!\! ds\,
\frac{s}{2\pi^3}\tanh\frac{\pi\Lambda}{s}
\int_{-i\infty}^{i\infty}\!\frac{dz}{2\pi i}e^{sz}\calF.
\label{eq:result_PsiQBcalGPsi}
\end{align}
It is natural that the $\delta$ dependence has disappeared
in \eqref{eq:result_PsiQBcalGPsi} since 
the dilatation property \eqref{eq:calGV3=V3} holds independently
of $\delta$, and so does the first equality of \eqref{eq:proof_one}.  
Eq.\,\eqref{eq:result_PsiQBcalGPsi} is independent of the order of the
$s$-integration and the limit $\Lambda\to \infty$ for the multi-brane
solutions since the region $s>\Lambda$ does not have a significant 
contribution to the $s$-integration.   

\begin{table}[htbp]
\begin{center}
\begin{tabular}{|c|c|c|c|c|c|}
\hline
$G(K)$ &$(n_0,n_\infty)$ & $\calN=\pi^2/3\int\Psi^3$
&$\pi^2\int \Psi\CR{\QB}{\calG}\Psi$ \\
\hline\hline
$K/(1+K)$ & $(1,0)$& $-1$ & $-1$ \\
\hline
$1/(1+K)$ & $(0,1)$& $-1$ & $-1$ \\
\hline
$1+1/K$ & $(-1,0)$& $1$ & $1$ \\
\hline
$1+K$ & $(0,-1)$& $1$ & $1$ \\
\hline
$K/(1+K)^2$ & $(1,1)$&$-2$ & $-2$ \\
\hline
$(1+K)^2/K$ & $(-1,-1)$& $2$ & $2$ \\
\hline
$K$ & $(1,-1)$& $0$ & $0$ \\
\hline
$1/K$ & $(-1,1)$& $0$ & $0$ \\
\hline
$(1+K)/(2+K)$& $(0,0)$ & $0$ & $0$\\
\hline
\end{tabular}
\end{center}
\caption{Comparison between $\calN$ and $\pi^2\int\! \Psi\CR{\QB}{\calG}\Psi$.}
\label{tab:Psi_QBcalG_Psi}
\end{table}
In table\,\ref{tab:Psi_QBcalG_Psi},
we show the values of \eqref{eq:result_PsiQBcalGPsi} 
for various $G(K)$.\footnote{
In this calculation, the regularization parameter $\eta$ for $K=\infty$
is unnecessary.
Namely, the result without $\eta$ from the start is 
the same as that with regularization. 
}
We also show the values of $\calN=\pi^2/3\int\! \Psi^3$ given by 
\eqref{eq:calN_n0_ninf} in terms of $(n_0,n_\infty)$ of 
\eqref{eq:n0_ninf}.
We see that the first equality of \eqref{eq:target}
certainly holds.
Therefore, \eqref{eq:result_PsiQBcalGPsi} is an inversion symmetric quantity
which can also count the singularity at $K=\infty$. 

\subsection{Extended gravitational coupling}
\label{sec:extend_gra}
In the previous subsection, we saw that \eqref{eq:Psi_QBcalG_Psi} agrees 
with $\calN$. The former consists of two terms, $\main$ and $\other$.
We find an important fact for the following arguments;
$\other$ for the multi-brane solutions vanishes 
by taking the limit $\delta \to 0$ after the $s$-integration.
We have checked this at least for $G(K)$ in 
table\,\ref{tab:Psi_QBcalG_Psi}.
Thanks to this fact, we have only to consider $\main$.

Here, we emphasize the importance of the regularization parameter
$\delta$ in obtaining our conclusion that $\calT_{\text{II}}$ may be
dropped. 
If we had put $\delta=0$ in $\calT_{\text{II}}$ from the start, we
would have met ill-defined quantities in various places; for example,
$\coth(2\pi y/s)$ in \eqref{eq:def_WIIB} is divergent 
at $y=\delta=0$. It is due to
the expression \eqref{eq:result_TII} for 
$\calT_{\text{II}}$ obtained by introducing
$\delta$ that we are allowed to argue that
$\calT_{\text{II}}$ can be discarded. Note also that, although
$\calT_{\text{I}}+\calT_{\text{II}}$ \eqref{eq:result_PsiQBcalGPsi} 
has turned out to be
independent of $\delta$, we have to keep $\delta$ when we consider
$\calT_{\text{I}}$ alone.

Then, let us consider $\calT_{\text{I}}$ which is expressed as 
\eqref{eq:T=2int_chiPsi}
in terms of the vertical integration $\gamma_L$ \eqref{eq:gamma_L}.
Using the EOM, we can rewrite it further as
\begin{align}
\main&=2\int \!\chL\Psi^2=2\int\!\chL\Gamma-2\int\! \ACR{\QB}{\chL}\Psi.
\label{eq:toward_V}
\end{align}
The first EOM term vanishes for the multi-brane solutions 
(see appendix \ref{sec:EOM_term}).
For the last term, we have  
\begin{align}
\ACR{\QB}{\chL}&=-\int_{\text{D}}^{\text{C}}\!\frac{dz}{2\pi i}
4\p\left(c\p X(z)c(\bar{z})\bar{\p}X(\bar{z})\right)
-\int_{\bar{\text{D}}}^{\bar{\text{C}}}\!\frac{d\bar{z}}{2\pi i}
4\bar{\p}\left(c\p X(z)c(\bar{z})\bar{\p}X(\bar{z})\right)\nn\\
&=\calV(D,\bar{\text{D}})-\calV(C,\bar{\text{C}}),
\label{eq:calN_calV_calV}
\end{align}
where $\calV(z,\bar{z})$ is the graviton emission vertex with zero momentum:  
\begin{align}
\calV(z,\bar{z})\equiv \frac{2}{\pi i}c\p X(z)c \bar{\p}X(\bar{z}).
\end{align}
The $\calV(C,\bar{\text{C}})=\Vmid$ part of \eqref{eq:toward_V} 
is nothing but the GIO \eqref{eq:def_GIO}. 
However, there is another term $\calV(D,\bar{\text{D}})=\Vend$ at the 
string endpoint. This is missing from the last term of \eqref{eq:target}.
Our expectation here is that this new term can count the singularity at
$K=\infty$ which the GIO could not.

\begin{figure}
\centering
\epsfxsize=0.40\textwidth
\epsfbox{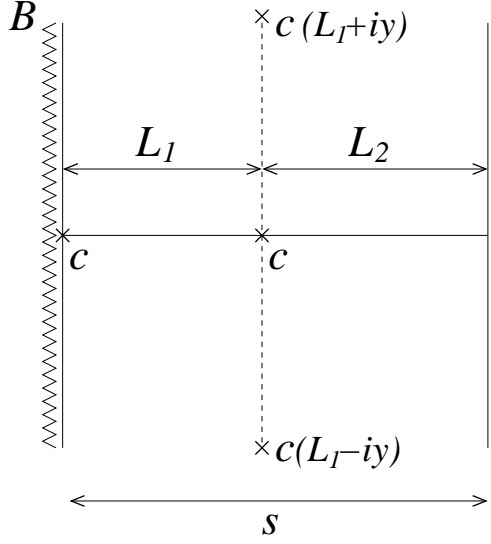}
\caption{Sliver frame for $\calI(y)$.
The vertical dashed line corresponds to the vertical paths CD and 
$\bar{\text{C}}\bar{\text{D}}$ in fig.\,\ref{fig:path_P}.}
\label{fig:calV_Psi}
\end{figure}

To confirm this expectation, 
let us evaluate $\int \ACR{\QB}{\chL}\Psi$. 
For this, we consider the following $\calI(y)$ and 
take the limits $y\to 0$ and $y\to \infty$:
\begin{align}
\calI(y)
&:=2\pi^2\int \calV(L_1+iy,L_1-iy)\Psi\nn\\
&=-4\pi i\int\left(\prod_{i=1}^2 dL_i\right)h(L_1)f(L_2)\nn\\
&\quad\times \VEV{\p X(L_1+iy)\bar{\p}X(L_1-iy)}_s
\VEV{Bc(0)c(L_1+iy)c(L_1-iy)c(L_1)}_s\nn\\
&=\frac{1}{2}\int\left(\prod_{i=1}^2 dL_i\right)h(L_1)f(L_2)
\left(\cosh\frac{\pi y}{s}\right)^{-2}
\left(y\sin \frac{2\pi L_1}{s}-L_1\sinh\frac{2\pi y}{s}\right)\nn\\
&=\int_0^\infty\!\! ds
\int_{-i\infty}^{i\infty}\!\frac{dz}{2\pi i}\,
e^{sz}\calD(s,z,y),
\label{eq:int_VyPsi}
\end{align}
with $\calD$ defined by 
\begin{align}
\calD(s,z,y)&=\frac{y}{4i}\left(\cosh\frac{\pi y}{s}\right)^{-2}(\Delta_s H)F
+\tanh\frac{\pi y}{s}H'F.
\label{eq:def_calD}
\end{align}
The sliver frame coordinate for \eqref{eq:int_VyPsi} 
is given in fig.\,\ref{fig:calV_Psi}.
Note that the first term of $\calD$ \eqref{eq:def_calD}
vanishes in both of the limits $y\to 0$ and $y\to \infty$.
And moreover, for the multi-brane solutions,
the contribution of this term to \eqref{eq:int_VyPsi} vanishes
when we take any of the two limits after the $s$-integration.
We have checked this at least for $G(K)$ 
in table\,\ref{tab:Psi_QBcalG_Psi}.\footnote{
For example, for $G=1+K$, we have $(\Delta_s H)F=(4\pi i/s)z/(1+z)$
and the contribution of the first term of $\calD$ to \eqref{eq:int_VyPsi}
is given by 
$$
\int_0^\infty\!\!ds\,\frac{\pi y}{s}
\left(\cosh\frac{\pi y}{s}\right)^{-2}\left(\delta(s)-e^{-s}\right),
$$
which vanishes in both the limits $y\to 0,\infty$.
Here, the $\delta(s)$ term has appeared since $(\Delta_s H)F$
is non-vanishing at $z=\infty$. 
The result remains unchanged even if we introduce the regularization parameters 
$\eps,\eta$ for $K=0,\infty$.
}
Therefore, $\calI(y)$ is reduced to  
\begin{align}
\calI(y)&=
-\int_0^\infty\!\!ds\,e^{-\eps s}
\tanh\frac{\pi y}{s} 
\int_{\eps -i\infty}^{\eps+i\infty}\!\frac{dz}{2\pi i}\,
e^{sz}\frac{zG'(z)}{G(z)}.
\label{eq:result_I}
\end{align}
Here, we have put $\eta=0$ in the integrand since the results are unchanged 
even if we take the limit $\eta \to 0$ after the integration.
In \eqref{eq:result_I} we have also redefined $z+\eps $ as $z$.

Let us calculate \eqref{eq:result_I} for $G(z)$
which is a rational function of $z$ and has no poles/zeros in $\text{Re}z>0$:
\begin{align}
G(z)=\prod_{i=0}^N\left(z+\alpha_i\right)^{n_i},\quad\left(\alpha_0=0,\
\text{Re}\,\alpha_i \ge 0\right).
\label{eq:rational_G}
\end{align}
Note that \eqref{eq:result_I} does not depend on the over all factor 
multiplying $G(z)$.
The index $n_0$ in \eqref{eq:n0_ninf} is equal to the present $n_{i=0}$, 
while $n_\infty$ in \eqref{eq:n0_ninf} is given by
\begin{align}
n_\infty=-n_0-\sum_{i=1}^N n_i.
\end{align}
For $G(z)$ \eqref{eq:rational_G}, we have 
\begin{align}
\frac{zG'(z)}{G(z)}=z\sum_{i=0}^N \frac{n_i}{z+\alpha_i}
=-n_\infty-\sum_{i=1}^N \frac{n_i\alpha_i}{z+\alpha_i}.
\label{eq:zG'/G}
\end{align}
For the last term of \eqref{eq:zG'/G}, 
the $z$-integration in \eqref{eq:result_I} 
is carried out by adding a large semicircle in the left half plane
(recall that $s>0$)
to consider the closed contour $C_L$, while the constant part 
$n_\infty$ gives the delta function:
\begin{align}
\int_{\eps-i\infty}^{\eps+i\infty}\!\frac{dz}{2\pi i}\,\, \frac{zG'}{G}e^{sz}
&=-n_{\infty}\delta(s)-\sum_{i=1}^N\int_{C_L} \!\frac{dz}{2\pi i}\,
\frac{n_i\alpha_i}{z+\alpha_i}e^{sz} 
=-n_{\infty}\delta(s)-\sum_{i=1}^N e^{-\alpha_i s}n_i\alpha_i.
\label{eq:ninfty+ni}
\end{align}
Carrying out the $s$-integration and taking the limits 
$y\to 0,\infty$, we obtain
\begin{align}
2\pi^2 \int \Vend \Psi&=\calI(y\to 0)=n_{\infty}, \label{eq:Vend}\\
2\pi^2\int \Vmid \Psi&=\calI(y\to \infty)
=n_{\infty}+\sum_{i=1}^N n_i=-n_0,
\label{eq:Vmid}
\end{align}
with $\Vend=\calV(D,\bar{\text{D}})$.
Here, the limits $y\to 0,\infty$ should be taken in principle 
after carrying out the $s$-integration. 
This is essential in evaluating the contribution 
of the $-n_\infty\delta(s)$ term of \eqref{eq:ninfty+ni} to $\calI(y\to 0)$.
However, the $s$-integration and the limits $y\to 0,\infty$ are 
exchangeable for the last term of \eqref{eq:ninfty+ni}. 
In particular, the exchange between the $s$-integration and the 
limit $y\to \infty$ is allowed for a pure imaginary $\alpha_i$
owing to the presence of $e^{-\eps s}$ in \eqref{eq:result_I}
(we take the limit $\eps\to 0$ after all the calculations).

\begin{figure}[tbp]
\centering
\epsfxsize=0.7\textwidth
\epsfbox{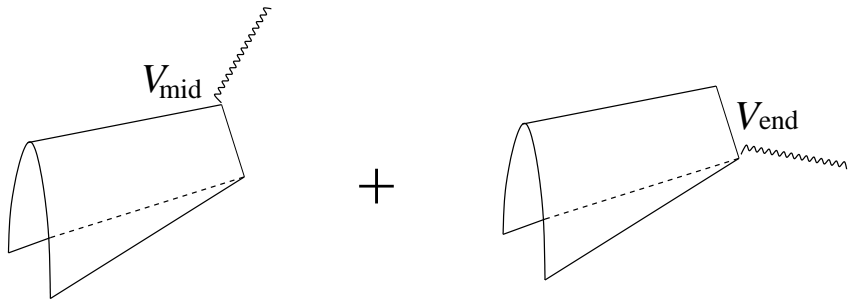}
\caption{Extended graviton coupling.
The graviton couples to the midpoint and the endpoint 
of the glued open string.}
\label{fig:Vmid_Vend}
\end{figure}

The result \eqref{eq:Vend} proves our expectation that
$\Vend$ can detect the singularity at $K=\infty$.
Finally, we obtain a inversion symmetric expression
\begin{align}
-2\pi^2\!\!\int\ACR{\QB}{\chL}\Psi&=
2\pi^2\left(\int \calV_{\text{mid}}\Psi-\int \calV_{\text{end}}\Psi\right)
=-n_0-n_\infty.
\label{eq:result_grav}
\end{align}
This agrees with $\calN$ given by \eqref{eq:calN_n0_ninf}
except the anomaly term $A(n_{0,\infty})$.
Namely, for the multi-brane solutions for which the EOM terms totally vanish,
the canonical energy $\calN$ is equal to extended graviton coupling:
\begin{align}
\frac{\calN}{2\pi^2}&=\frac{1}{2}\int \Psi*\CR{\QB}{\calG}\Psi=
\int \Vmid \Psi-\int\Vend \Psi.
\label{eq:improve_N=GIO}
\end{align}
This is the correct relation which should take the place of \eqref{eq:target}.
The graviton couples to both the midpoint and the endpoint of open string
(see fig.\,\ref{fig:Vmid_Vend}). 

Some comments are in order concerning our results.
First, the same equation as \eqref{eq:result_I} without $\tanh(\pi y/s)$
appears in \cite{MS2} in their analysis of the GIO, namely, $\calI(y\to \infty)$
(see eq.\,(2.36) in \cite{MS2}).
Indeed, $\tanh(\pi y/s)$ may be set equal to $1$ before the $s$-integration 
for the GIO.
However, as stated above, the presence of $\tanh(\pi y/s)$ is indispensable
in order for $\calI(y\to 0)$ to be able to detect the $K=\infty$ singularity. 

Our second comment is on the treatment of the $-n_{\infty}\delta(s)$ term
of \eqref{eq:ninfty+ni}.
In obtaining \eqref{eq:Vend} and \eqref{eq:Vmid},
we used $\int_{0}^\infty\! ds\, \delta(s)\tanh(\pi y/s)=1$ for the 
delta function $\delta(s)=\int_{-i\infty}^{i\infty}dz/(2\pi i) e^{sz}$.
However, if we adopt $\int_{0}^\infty\! ds\, \delta(s)\tanh(\pi y/s)=1/2$
by taking into account that the $s$-integration is only for $s>0$,
we obtain results different from \eqref{eq:Vend} and \eqref{eq:Vmid}:
$\calI(y\to 0)=n_{\infty}/2$ and $\calI(y\to \infty)=-n_{0}-n_{\infty}/2$.
Furthermore, if we start the $sz$-trick by inserting 
$1=\int_{-\infty}^{\infty}\!ds\,\delta(s-\sum_{i}L_i)$ with the range of 
the $s$-integration extended to $(-\infty,\infty)$,
and use $\int_{-\infty}^{\infty}\!ds\,\delta(s)\tanh(\pi y/s)=0$,
we obtain the result that the GIO can count the singularities both at 
$K=0$ and $K=\infty$:
$\calI(y\to 0)=0$ and $\calI(y\to \infty)=-n_{0}-n_{\infty}$.
These facts show that the $sz$-trick of inserting 
$1=\int \!ds\, \delta(s-\sum_i L_i)$ and exchanging the order of the 
$L_i$ and $s$ integrations is not well-defined when 
there appears $\delta(s)$ after the $z$-integration.
However, fortunately, the difference $\calI(y\to \infty)-\calI(y\to 0)$ 
is free from the subtleties and is always equal to $-n_0-n_{\infty}$.
In fact, the $\delta(s)$ term vanishes in the difference since 
we have $\left(\tanh(\pi \Lambda/s)-\tanh(\pi \delta/s)\right)\delta(s)=0$.
Namely, it does not make sense to discuss the values of respective quantities,
$\int \Vmid\Psi$ and $\int \Vend\Psi$. 
Only their difference is of significance. 

Thirdly, we reemphasize the point for the the second equality of 
\eqref{eq:improve_N=GIO} relating \eqref{eq:Psi_QBcalG_Psi} 
quadratic in $\Psi$ to another expression linear in
it. As we mentioned below \eqref{eq:QBchi=V}, 
in order for this manipulation by use of
the EOM, $\Psi^2=-\QB\Psi$, to be possible, it is necessary that
$\CR{\QB}{\calG}$ is effectively reduced to the sum of the two terms
on the left and the right vertical paths, CD and AB,
respectively. Namely, the contribution of the horizontal path part in
$\CR{\QB}{\calG}$ should vanish in the limits $z_0\to\infty$,
$\Lambda\to\infty$ and $\delta\to 0$.
As we saw in Secs.\,\ref{sec:eval_T_I} and \ref{sec:Evalu_T_II} 
this is in fact realized; firstly,
in $\calC_\text{I}$ in \eqref{eq:CR_QB_calG}, only its vertical path part 
survives in the limit $\Lambda\to\infty$, and secondly,
the sum of the contributions to \eqref{eq:Psi_QBcalG_Psi} 
of the $\calC_{\text{IIA}}$ and
$\calC_{\text{IIB}}$ terms in \eqref{eq:CR_QB_calG}, namely  $\calT_{\text{II}}$,
vanishes in the limit $z_0\to \infty$ and $\delta\to 0$.

\subsection{Relation to the argument in \cite{BI}}
\label{sec:relate_BI}
Finally in this section,
we mention the relationship between our argument 
leading to \eqref{eq:improve_N=GIO} and that of \cite{BI} giving 
\eqref{eq:target} which contains the GIO alone.
The discrepancy stems from different ways of taking the limit
$\delta\to 0$ for \eqref{eq:Psi_QBcalG_Psi}.
In our analysis, we take the limit $\delta\to 0$ after carrying out
the $s$-integration for \eqref{eq:Psi_QBcalG_Psi}. We saw above that
this is essential for detecting the singularity at $K=\infty$.
On the other hand, in \cite{BI}, they estimated the behavior of the
operators in $\CR{\QB}{\calG}$ for small $\delta$ before carrying out
the $s$-integration and, in addition, without taking into account the
presence of $\Psi$.
As a result, they argued that, among the terms in $\calT_{\text{II}}$,
only $\kappa(D,\bar{\text{D}})$ in $\calT_{\text{IIB}}$
\eqref{eq:surface_term} needs to be kept
since it contains the $1/\delta$ singularity due to $\VEV{X \p X}$.
Namely, their $\chi$ (i.e. $\chi_L$ is
Sec.\,\ref{sec:relation_calN_calV}) given in page 11 of \cite{BI} is
related to $\chL$ by
\begin{align}
\chi_L=\chL-\kappa(D,\bar{\text{D}}),
\label{eq:chi=gamma-k}
\end{align}
with
$\kappa(D,\bar{\text{D}})\sim
-(c(\text{D})+c(\bar{\text{D}}))/(4\pi\delta)$. 
Since we have
$\ACR{\QB}{\kappa(\text{D},\bar{\text{D}})}=\Vend+O(\delta)$,
two $\Vend$'s in $\ACR{\QB}{\chi_L}$ cancel to give
\eqref{eq:QBchi=V} and hence \eqref{eq:target}.
This argument without taking into account the effects of $\Psi$ in
\eqref{eq:Psi_QBcalG_Psi} applies to both of the cases of hermitian
$\Psi$ adopted in \cite{BI} and non-hermitian one
\eqref{eq:general_Psi} used in this paper.\footnote{
There arises a special feature for non-hermitian $\Psi$
\eqref{eq:general_Psi} if we take into account the presence of $\Psi$
in \eqref{eq:Psi_QBcalG_Psi};
the contribution of $\kappa(\text{D},\bar{\text{D}})$ vanishes by
itself in the naive limit $\delta\to 0$ with $s>0$ fixed.
This is understood from \eqref{eq:result_calWIIB} and also from
the fact that our $\Psi$ \eqref{eq:general_Psi} has $c$ on the left
which collides with $c$ in $\kappa(D,\bar{\text{D}})$ in the limit
$\delta\to 0$.
However, \eqref{eq:target} holds in any case since
the contribution of $\Vend$ in $\ACR{\QB}{\chL}$ also vanishes
in this naive $\delta\to 0$ limit as we saw
in Sec.\,\ref{sec:extend_gra}.
}

Our concrete calculations in this section for non-hermitian
$\Psi$ \eqref{eq:general_Psi} show the followings.
Keeping the $\kappa(\text{D},\bar{\text{D}})$ term alone corresponds to 
taking only the part of $\calW_{\text{IIB}}$ \eqref{eq:def_WIIB} 
multiplied by $(1/s)\coth (2\pi \delta/s)\sim 1/\delta$ $(\delta\to 0)$ 
and throwing away 
all other contributions to $\calT_{\text{II}}$.
However, this cannot be accepted in our analysis where we take
the limit $\delta\to 0$ after the $s$-integration.
In fact, the contribution of the $\kappa(\text{D},\bar{\text{D}})$
term including the $KBc$ correlator part (namely, the presence of $\Psi$) 
is given by the part of 
$\calW_{\text{IIB}}$ \eqref{eq:result_calWIIB} multiplied by
$\cosh(2\pi\delta/s)\tanh(\pi\delta/s)$.
This leads to a divergent $s$-integration at $s=0$ for a finite
$\delta$.
Fortunately, all the terms in $\calT_{\text{II}}$ are equally
important when we take into account the presence of the $KBc$
correlator part as seen from \eqref{eq:result_calWIIA} and
\eqref{eq:result_calWIIB}, and their sum gives \eqref{eq:result_TII}
with regular integrand at $s=0$.
It would be desirable if we could repeat concrete calculations in this
section for hermitian $\Psi$. However, this seems technically a
difficult task.

\section{Conclusion}
\label{sec:conclusion}
In this paper, we examined the equivalence between 
the canonical energy $\calN$ and the gravitational coupling 
for the multi-brane solutions of pure-gauge type, 
$\Psi=U\QB U^{-1}$ with $U$ given by \eqref{eq:U_general_form}
in terms of the function $G(K)$.
Especially, we considered only 
$\Psi$ satisfying the EOM in the strong sense, and therefore  
restricted ourselves to $G(K)$ with $n_0,n_\infty=0,\pm 1$
($n_0$ and $n_\infty$ are defined by \eqref{eq:n0_ninf}).
Although, singularities at $K=0$ and $K=\infty$ 
carry out the same work in $\calN$ (inversion symmetry), 
the GIO, which has been regarded as the gravitational coupling of open string, 
cannot detect the singularity at $K=\infty$.
Therefore, it is a problem that, for the solutions carrying the singularity at 
$K=\infty$, the GIO is different from the canonical energy.
In \cite{BI}, however, a proof was given which shows the equality of 
the canonical energy and the GIO for a class of multi-brane solutions
we are considering.
For resolving this contradiction, we examined every step in the proof
of \cite{BI} concretely for the multi-brane solutions.
As a result, we found that the correct gravitational coupling 
consists of the sum of the GIO and another new term $\int\Vend \Psi$.
Compared with the GIO, $\int\Vmid\Psi$ with the graviton emission vertex 
at the string midpoint, the new term contains the endpoint insertion.
We found that the sum $\int(\Vmid-\Vend)\Psi$ is a well-defined quantity 
free from the ambiguity in its $sz$-integration for $G(K)$ 
with non-vanishing $n_{\infty}$.
And this sum is indeed a inversion symmetric quantity which can detect
the singularity at $K=\infty$ as well as that at $K=0$. 

Finally, we comment on the gauge invariance of the gravitational
coupling. Indeed, the GIO $\int\!\calV_{\text{mid}}\Psi$ is exactly
invariant under any infinitesimal gauge transformation
\begin{align}
\delta_\lambda\Psi=\QB\lambda+\Psi*\lambda-\lambda*\Psi,
\label{eq:gauge_trans}
\end{align}
and for any $\Psi$ not restricted to solutions to the EOM.
This is owing to the fact that the graviton emission vertex is
inserted at the string midpoint in the GIO \cite{HI,GRSZ,Ell}.
On the other hand, $\int\!\calV_{\text{end}}\Psi$ is not gauge
invariant except for $\Psi$ and $\lambda$ which are inert under
the twist transformation $\sigma\to\pi-\sigma$.
Namely, $\int\!\calV_{\text{end}}\Psi$ alone cannot have a universal
physical meaning.
It is true that, for the multi-brane solutions we are considering,
the canonical energy is equal to the sum of the
gravitational couplings at the string midpoint and at the endpoint as
given in \eqref{eq:improve_N=GIO}. 
However, this would not necessarily be the case for
other classes of solutions.

Then, where have we lost the gauge invariance?
In \eqref{eq:improve_N=GIO}, the starting quantity 
$\calN=(\pi^2/3)\int\!\Psi^3$
and the second expression $\int\Psi\CR{\QB}{\calG}\Psi$ are gauge
invariant up to the EOM.
In obtaining the last expression of \eqref{eq:improve_N=GIO},
$\int\!\left(\calV_{\text{mid}}-\calV_{\text{end}}\right)\Psi$,
we discarded two quantities.
One is the horizontal integration part in $\main$
and the other is $\other$.
The former vanishes in the limit $\Lambda\to \infty$ and the 
latter in the limit $z_0\to \infty$ and $\delta\to 0$ 
for our multi-brane solutions.
However, these two quantities are not gauge invariant even under the use
of the EOM.
Thus we lost the invariance of the final
expression of \eqref{eq:improve_N=GIO}
under the gauge transformation 
\eqref{eq:gauge_trans} for generic $\Psi$ and $\lambda$.

Although the generic gauge invariance is lost in
$\int\!\left(\calV_{\text{mid}}-\calV_{\text{end}}\right)\Psi$,
our result \eqref{eq:result_grav} implies that it is invariant 
under the gauge transformations which do
not change $(n_0,n_\infty)$ specifying the singularity structure at
$K=0$ and $\infty$.
(Note that the deformation of $G(K)$, namely, that of $U$ is a 
gauge transformation \eqref{eq:gauge_trans} with $\lambda=-\delta U U^{-1}$.) 
It is desirable to understand the mechanism of invariance 
under this special type of gauge transformation.\footnote{
Even if we adopt hermitian (twist-even) $\Psi$ with
$U=1-\sqrt{1-G}Bc\sqrt{1-G}$ instead of \eqref{eq:U_general_form},
$\lambda=-\delta U\,U^{-1}$ is not necessarily twist-even.
Therefore, the gauge invariance of $\int\!\calV_{\text{end}}\Psi$
for twist-even $\Psi$ and $\lambda$ cannot be used for the present
purpose.}

\section*{Acknowledgments}
We would like to thank T.~Baba, N.~Ishibashi and T.~Masuda for valuable
discussions.
The work of H.~H.\ was supported in part by a Grant-in-Aid for
Scientific Research (C) No.~25400253 from JSPS.
The work of T.~K.\ was supported in part by a Grant-in-Aid for
JSPS Fellows No.~24$\cdot$1601.

\appendix
\section{Correlators in the sliver frame}
\label{sec:correlator}
In this appendix, we summarize the correlators on the sliver frame 
with width $s$:
\begin{align}
\VEV{\p X(z) \p X(z')}_s&=\frac{\pi^2}{2s^2}
\left(\sin\frac{\pi (z-z')}{s}\right)^{-2},
\label{eq:pXpX}\\
\VEV{\p X(z)\bar{\p}X(\bar{z}')}_s&=
\frac{\pi^2}{2s^2}\left(\sin\frac{\pi(z-\bar{z}')}{s}\right)^{-2},
\label{eq:pXbarpX}\\
\VEV{X(z,\bar{z})\p X(z')}_s&=-\frac{\pi}{2s}\left(
\cot\frac{\pi(z-z')}{s}+\cot\frac{\pi(\bar{z}-z')}{s}
\right),\\
\VEV{X(z,\bar{z})\bar{\p}X(\bar{z}')}_s&=
-\frac{\pi}{2s}\left(
\cot\frac{\pi(z-\bar{z}')}{s}+\cot\frac{\pi(\bar{z}-\bar{z}')}{s}
\right),\\
\VEV{c(z_1)c(z_2)c(z_3)}_s&=\left(\frac{s}{\pi}\right)^3
\sin\frac{\pi(z_1-z_2)}{s}\sin\frac{\pi(z_2-z_3)}{s}\sin\frac{\pi(z_1-z_3)}{s}
,\\
\VEV{Bc(z_1)c(z_2)c(z_3)c(z_4)}_s
&=\frac{s^2}{\pi ^3} \left\{z_4 \sin \frac{\pi 
   (z_1-z_2)}{s} \sin\frac{\pi 
   (z_1-z_3)}{s}\sin\frac{\pi 
   (z_2-z_3)}{s}\right.\\
&\left.\quad-z_1 \sin \frac{\pi 
   (z_2-z_3)}{s} \sin \frac{\pi 
   (z_2-z_4)}{s} \sin \frac{\pi 
   (z_3-z_4)}{s}\right.\\
&\left.\quad-z_3 \sin\frac{\pi 
   (z_1-z_2)}{s} \sin \frac{\pi 
   (z_1-z_4)}{s} \sin \frac{\pi 
   (z_2-z_4)}{s}\right.\\
&\left.\quad+z_2 \sin \frac{\pi 
   (z_1-z_3)}{s}\sin \frac{\pi 
   (z_1-z_4)}{s} \sin \frac{\pi 
   (z_3-z_4)}{s}\right\}.
\label{eq:Bcccc}
\end{align}
From \eqref{eq:Bcccc}, we get immediately,
\begin{align}
&2\,\text{Re}\VEV{Bc(0)c(L_1+iy)c(L_1)c(L_1+L_2)}_s
=\sum_{\pm}\VEV{Bc(0)c(L_1\pm iy)c(L_1)c(L_1+L_2)}_s\nn\\
&=-\frac{s^2}{\pi ^3} \left(\sinh \frac{\pi  y}{s}\right)^2 \left(L_2 \sin
   \frac{2 \pi  L_1}{s}-L_1 \sin \frac{2
   \pi  L_2}{s}\right),
\label{eq:real_Bcccc}
\end{align}
and
\begin{align}
\VEV{Bc(0)c(L_1+iy)c(L_1-iy)c(L_1)}_s=\frac{is^2}{\pi^3}
\left(\sinh\frac{\pi y}{s}\right)^2\left(
L_1\sinh \frac{2\pi y}{s}-y \sin\frac{2\pi L_1}{s}\right).
\end{align}
Finally, we present the expectation values of $g_z$ and $g_{\bar{z}}$ 
defined by \eqref{eq:def_g} and \eqref{eq:normalordering}:
\begin{align}
&\VEV{g_z(z,\bar{z})}_s
=\frac{\pi}{s}\left[\cot\frac{\pi(z-\bar{z})}{s}
+\cot\frac{\pi(z_0-z)}{s}+\cot\frac{\pi(\bar{z}_0-z)}{s}
\right],\nn\\
&\VEV{g_{\bar{z}}(z,\bar{z})}_s
=-\frac{\pi}{s}\left[\cot\frac{\pi(z-\bar{z})}{s}
-\cot\frac{\pi(z_0-\bar{z})}{s}-\cot\frac{\pi(\bar{z}_0-\bar{z})}{s}
\right].\label{eq:VEV_g_z_g_bz}
\end{align}

\section{Calculation of the EOM-terms}
\label{sec:EOM_term}
Our result in this paper including the EOM-terms is given by 
\begin{align}
\frac{\calN}{2\pi^2}&=\int\left(\Vmid-\Vend\right)\Psi
+\frac{1}{2}\int \Psi*\Gamma
-\int(\calG\Psi)*\Gamma
+\int\chL\Gamma,
\label{eq:app_N_V}
\end{align}
which should replace \eqref{eq:proof_five}.
In the EOM-terms containing $\Gamma$,
$\Psi$ denotes the regularized one $\Psi_{\eps\eta}$, 
and $\Gamma$ is given by \eqref{eq:def_Gamma}.
We know already that the second term
on the RHS of \eqref{eq:app_N_V} vanishes for our multi-brane solutions 
(see \eqref{eq:EOM_test}).
In this appendix, we show that the other EOM-terms also vanish.
In \cite{BI}, the same kind of calculation was carried out for 
$G(K)$ with $n_\infty=0$ ($\chL$ is replaced with $\chi_L$ in \cite{BI}).
The difference between the calculation in \cite{BI} and ours 
is that we take the limits $\Lambda\to \infty$ and $\delta\to 0$
after the $s$-integration, while the order is exchanged in \cite{BI}.
However, the result of \cite{BI} and ours are consistent for $G(K)$
with $n_{\infty}=0$. 

Explicitly, $\Gamma$ \eqref{eq:def_Gamma} is given as the sum of two terms:
\begin{align}
\Gamma=\QB\Psi_{\eps\eta}+\Psi_{\eps\eta}^2=\eps\times 
\Gamma^{(\eps)}[G(K_{\eps\eta})]
+\eta \times \Gamma^{(\eta)}[G(K_{\eps\eta})],
\end{align}
where $\Gamma^{(\eps)}$ and $\Gamma^{(\eta)}$ are defined by
\begin{align}
\Gamma^{(\eps)}[G(K_{\eps\eta})]&:=cFcH,\nn\\
\Gamma^{(\eta)}[G(K_{\eps\eta})]&:=
cK_\eps^2\CR{\frac{F}{K_\eps^2}}{c}K_\eps^2 BcH,\nn
\end{align}
with $F$ and $H$ given in \eqref{eq:def_FHJ}.
In order for the $\Gamma^{(\eps)}$ part to be non-vanishing in the limit
$\eps\to 0$, it should behave as $O(1/\eps)$.
Such singularity is expected to arise from $G(K)$ with $n_0\neq 0$.
Likewise, $G(K)$ with $n_{\infty}\neq 0$ could make the $\Gamma^{(\eta)}$
part non-trivial in the limit $\eta\to 0$.
However, in the rest of this appendix,
we examine both of the two terms $\Gamma^{(\eps)}$ and $\Gamma^{(\eta)}$ for 
a given $G(K)$.

\subsection{$\int \chL\Gamma$}
First we consider $\int\chL\Gamma$.
We obtain after tedious calculations
\begin{align}
\int \chL \Gamma^{(\eps)}&=
\frac{1}{8\pi^3 i}
\int \! ds \, s^2 \left[\tanh\frac{\pi y}{s}\right]^{y=\Lambda}_{y=\delta}
\int_{-i\infty}^{i\infty}\! \frac{dz}{2\pi i} e^{sz} (\Delta_s H)F,\nn\\
\int \chL \Gamma^{(\eta)}&=\frac{1}{8\pi^3 i}\int\! ds \, s
\left[\tanh\frac{\pi y}{s}\right]^{y=\Lambda}_{y=\delta}\nn\\
&\quad \times \int_{-i\infty}^{i\infty}\!\frac{dz}{2\pi i}e^{sz}
\left\{
(-(\Delta_s H)F'+H'(\Delta_sF))z_{\eps}^2-
(F\leftrightarrow z_{\eps}^2)
\right\},\label{eq:chLGamma}
\end{align}
where $\Delta_s$ is defined by \eqref{eq:def_Delta_s} and $z_\eps:=z+\eps$.
In table \ref{tab:chLGamma}, we show the result of calculations of 
the two quantities of
\eqref{eq:chLGamma} for various $G(K)$, which essentially
cover all possible $G(K)$ for non-trivial multi-brane solutions
with $\calN=\pm 1,\pm 2$. 
Since we take the limit $\delta\to 0$ and $\Lambda\to \infty$
after the $s$-integration,
our analyses are mostly based on numerical $s$ integrations.  
As shown in the table, the two quantities are of $O(1)$ in the limit
$\eps,\eta\to 0$ in most cases. Exceptions are $\int\chL\Gamma^{(\eps)}$ 
for $G=1+1/K$ and $(1+K)^2/K$. In these cases, there appears $O(\ln \eps)$
which is due to the poles of the $z$ integrand on the imaginary axis.
In any case, $\int\chL \Gamma$ vanishes in the limit $\eps,\eta\to 0$.  
\begin{table}[htbp]
\begin{center}
\begin{tabular}{|l|l|l|l|}
\hline
$G(K)$ & $(n_0,n_{\infty})$ & $\int \chL\Gamma^{(\eps)}$ 
& $\int \chL\Gamma^{(\eta)}$\\
\hline\hline
$K/(1+K)$ & $(1,0)$ & $O(1)$ & $O(1)$ \\
\hline
$1/(1+K)$ & $(0,1)$ & $O(1)$ & $O(1)$\\
\hline
$1+1/K$ & $(-1,0)$ & $O(\ln \eps)$ & $O(1)$\\
\hline
$1+K$ & $(0,-1)$ & $O(1)$ & $O(1)$ \\
\hline
$K/(1+K)^2$ & $(1,1)$ & $O(1)$ & $O(1)$\\
\hline
$(1+K)^2/K$ & $(-1,-1)$ & $O(\ln \eps)$ & $O(1)$ \\
\hline
\end{tabular}
\end{center}
\caption{$\int \chL\Gamma^{(\eps)}$ and $\int \chL\Gamma^{(\eta)}$
for various $G(K)$.}
\label{tab:chLGamma}
\end{table}

\subsection{$\int (\calG\Psi)*\Gamma$}
Next, let us consider $\int(\calG\Psi)*\Gamma$.
Since $\Psi$ and $\Gamma$ does not contain $X(z,\bar{z})$ explicitly,
this quantity factorizes into the product of $\VEV{\calG}_s$ and 
the $KBc$ correlators. 
The $(z_0,\bar{z}_0)$ part in $\VEV{\calG}_s$ vanishes in the 
limit $z_0\to \infty$, and 
the two vertical integration parts in the rest of $\VEV{\calG}_s$ cancel
as seen from \eqref{eq:VEV_g_z_g_bz}.
Therefore, we are left with the horizontal integration part given by 
\eqref{eq:VEV_calG} with $L_i$ being the width of $\Psi$.
Using these facts, we obtain 
\begin{align}
\int (\calG\Psi)*\Gamma^{(\eps)}&=
\langle\!\langle H',F,H,F\rangle\!\rangle
+\langle\!\langle H,F,H,F'\rangle\!\rangle,\nn\\
\int (\calG\Psi)*\Gamma^{(\eta)}&=
\langle\!\langle HF',H,F,K_{\eps}^2\rangle\!\rangle
+\langle\!\langle HF,H',F,K_{\eps}^2\rangle\!\rangle\nn\\
&\quad-\langle\!\langle F',H,F,K_{\eps}^2H\rangle\!\rangle
-\langle\!\langle F,H',F,K_{\eps}^2H\rangle\!\rangle\nn\\
&\quad-\left\{
\langle\!\langle HF',H,K_{\eps}^2,F\rangle\!\rangle
+\langle\!\langle HF,H',K_{\eps}^2,F\rangle\!\rangle\right.\nn\\
&\left.\quad
-\langle\!\langle F',H,K_{\eps}^2,FH\rangle\!\rangle
-\langle\!\langle F,H',K_{\eps}^2,FH\rangle\!\rangle\
\right\},
\end{align}
with
\begin{align}
\langle\!\langle F_1,F_2,F_3,F_4 \rangle\!\rangle&:=
-\frac{1}{(2\pi)^3 i}
\int_0^\infty\!\! ds\int_{-i\infty}^{i\infty}\frac{dz}{2\pi i}\,e^{sz}\, 
s\coth\frac{2\pi\Lambda}{s}\nn\\
&\quad\times \left[
(\Delta_s F_1)F_2F_3'+F_1'F_2(\Delta_s F_3)+(F_1(\Delta_s F_2)F_3)'
-\Delta_s(F_1F_2')F_3\right.\nn\\
&\left.\quad-\Delta_s(F_1F_2)F_3'-F_1'\Delta_s(F_2F_3)
-F_1\Delta_s(F_2'F_3)+\Delta_s (F_1F_2'F_3)
\right]F_4.
\label{eq:sz_formula}
\end{align}
In \eqref{eq:sz_formula}, $F_i(K)$ on the LHS should be replaced with 
$F_i(z)$ on the RHS.
Table \ref{tab:gPsi_Gamma} shows the result 
of our calculation for various $G(K)$.
We see that $\int(\calG\Psi)*\Gamma$ vanishes in the limits $\eps,\eta\to 0$
for all the multi-brane solutions.
\begin{table}[hbtp]
\begin{center}
\begin{tabular}{|l|l|l|l|}
\hline
$G(K)$ & $(n_0,n_\infty)$ & $\int \calG\Psi*\Gamma^{(\eps)}$ 
& $\int \calG\Psi*\Gamma^{(\eta)}$\\
\hline
\hline
$K/(1+K)$ & $(1,0)$ & $O(1)$ & $O(1)$ \\
\hline
$1/(1+K)$ & $(0,1)$ & $O(1)$ & $O(1)$\\
\hline
$1+1/K$ & $(-1,0)$ & $O(\ln \eps)$ & $O(1)$\\
\hline
$1+K$ & $(0,-1)$ & $O(1)$ & $O(1)$ \\
\hline
$K/(1+K)^2$ & $(1,1)$ & $O(1)$ & $O(1)$\\
\hline
$(1+K)^2/K$ & $(-1,-1)$ & $O(\ln\eps)$ & $O(1)$ \\
\hline
\end{tabular}
\end{center}
\caption{$\int \calG\Psi*\Gamma^{(\eps)}$ and $\int \calG\Psi*\Gamma^{(\eta)}$
for various $G(K)$.}
\label{tab:gPsi_Gamma}
\end{table}
\section{BRST Ward-Takahashi identity}
\label{sec:BRST_W_T_identity}
Recall that $g_z$ and $g_{\bar{z}}$ \eqref{eq:def_g} contain $X(z_0,\bar{z}_{0})$.
This part made a non-trivial contribution to $\calT_{\text{II}}$ in the limit 
$z_0\to \infty$.
The original reason why this term was introduced is for the validity of 
the BRST Ward-Takahashi identity (WTI), which was intensively used in 
the derivation of \eqref{eq:improve_N=GIO}.
In this appendix, we see explicitly how the $X(z_0,\bar{z}_0)$
term is needed for the BRST WTI.

Let us consider the following quantity:  
\begin{align}
\VEV{\QB\left\{\calM(z,\bar{z};z'\bar{z}')Bc(z_1)c(z_2)c(z_3)\right\}}_s,
\label{eq:QB_calM_Bccc}
\end{align}
where $\calM(z,\bar{z};z',\bar{z}')$ is an operator consisting only 
of $X$s and their derivatives at $(z,\bar{z})$ and $(z',\bar{z}')$.   
The vanishing of \eqref{eq:QB_calM_Bccc} is an example of the BRST WTI.

\eqref{eq:QB_calM_Bccc} consists of two terms in which 
$\QB$ acts on $\calM$ and the $Bccc$ part, respectively.
Using that $K=\ACR{\QB}{B}$ is equal to $-\p /\p s$ acting on both 
$\VEV{\calM}_s$ and $\VEV{ccc}_s$,
and that the BRST WTI $\VEV{\QB (Bccc)}=0$ 
for a purely ghost quantity holds valid,
we obtain
\begin{align}
\eqref{eq:QB_calM_Bccc}
&=\VEV{\left[\QB\calM(z,\bar{z};z',\bar{z}')\right]Bc(z_1)c(z_2)c(z_3)}_s
-\left(\p_s\VEV{\calM}_s\right)\VEV{c(z_1)c(z_2)c(z_3)}_s.
\end{align}
Especially, choosing $\calM=X(z,\bar{z})\p X(z')$ in \eqref{eq:QB_calM_Bccc}, 
we get a non-vanishing quantity:
\begin{align}
-\frac{1}{2\pi }\left\{(z_2-z_3) \sin \frac{2 \pi 
   (z_1-z')}{s}+(z_3-z_1) \sin\frac{2 \pi 
   (z_2-z')}{s}+(z_1-z_2) \sin \frac{2 \pi 
   (z_3-z')}{s}\right\}.
\label{eq:calA}
\end{align}
Namely, the BRST WTI does not hold when $\calM$ contains $X$ explicitly.
However, since \eqref{eq:calA} is independent of $(z,\bar{z})$, 
we see that the BRST WTI holds for  
$\calM=\p X(z)\p X(z')$ and $\bar{\p}X(\bar{z})\p X(z')$. 
The BRST WTI also holds for 
$\calM=\left(X(z,\bar{z})-X(z_0,\bar{z}_0)\right)\p X(z')$:
\begin{align}
\VEV{\QB\left\{\left(X(z,\bar{z})-X(z_0,\bar{z}_0)\right)\p X(z')
Bc(t_1)c(t_2)c(t_3)\right\}}=0.
\end{align}
This is the identity we used in the derivation of \eqref{eq:improve_N=GIO}.

\end{document}